# Spin-1 photons, spin-½ electrons, Bell's inequalities, and Feynman's special perspective on quantum mechanics

Masud Mansuripur
James C. Wyant College of Optical Sciences, The University of Arizona, Tucson

**Abstract**. The Einstein-Podolsky-Rosen (EPR) paradox that argues for the incompleteness of quantum mechanics as a description of physical reality has been put to rest by John Bell's famous theorem, which inspired numerous experimental tests and brought about further affirmations of quantum reality. Nevertheless, in his writings and public presentations, Richard Feynman never acknowledged the significance of Bell's contribution to the resolution of the EPR paradox. In this paper, we discuss several variants of the Bell inequalities (including one that was specifically espoused by Feynman), and explore the ways in which they demolish the arguments in favor of local hidden-variable theories. We also examine the roots of Feynman's attitude toward Bell's theorem in the context of Feynman's special perspective on quantum mechanics.

**1. Introduction**. Quantum weirdness is rooted in the fact that the states of a quantum system have probability amplitudes, that these amplitudes are complex numbers, and that the evolution, superposition, and combination of these complex amplitudes give rise to phenomena that often seem strange, counterintuitive, and outright paradoxical to the worldview that is firmly rooted in our understanding of classical physics, where probability amplitudes play no role. Richard Feynman has consistently and persistently emphasized the importance of probability amplitudes in his teachings and his descriptions of quantum phenomena.[1,2] In fact, his path integral formulation of quantum mechanics is based on the fundamental idea that one must add up all the probability amplitudes that contribute to an event along different paths — so long as the paths are physically allowed and are, in principle, indistinguishable from one another.[3] The squared magnitude of the sum total of these complex amplitudes will then yield the ordinary probability of occurrence of the event.[1-3]

Given Feynman's profound understanding of physics, in general, and of quantum mechanics, in particular, one would be inclined to think that he had reached the point where quantum weirdness no longer disturbed him. Yet, as late as 1982, he would write:[4] "*we always have had a great deal of difficulty in understanding the world view that quantum mechanics represents. At least I do, because I'm an old enough man that I haven't got to the point that this stuff is obvious to me. Okay, I still get nervous with it.*" In the same paper,[4] Feynman describes an experiment with a pair of polarization entangled photons which demonstrates the incompatibility of quantum mechanical predictions with the existence of local hidden variables. While the argument is very much in line with John Bell's views on the subject,[5-7] Feynman does not credit Bell, and indeed appears to have quietly known and internalized the implications of such local hidden-variable theories (LHVTs) for quite some time.[8,9] Be it as it may, the goal of the present paper is *not* to adjudicate the issue for purposes of parceling out historical credit, but rather to provide an accessible account of Bell's inequalities that is informed by Feynman's special perspective on quantum mechanics — specifically, his belief in the primacy of probability amplitudes.[1-4]

The organization of the paper is as follows. The maximally entangled photon pair discussed by Feynman in the aforementioned paper[4] is the subject of Sec.2. Here, we begin by describing the polarization state of the photon pair, then present the quantum mechanical explanation of experiments in which each photon passes through a linear polarizer (also known as polarization analyzer) before being detected. This is followed by a concise argument as to why any LHVT cannot possibly explain the observed correlations in certain experiments of the sort. In the particular arrangement of polarizers chosen by Feynman, the probability of agreement between the signals of the two detectors (i.e., both receive a photon or neither does) is ¾ according to the quantum theory, but cannot exceed ⅔ if an LHVT happens to be operative. It is in the context of this observation that Feynman concludes by saying: "*I have entertained myself always by squeezing the difficulty of quantum mechanics into a smaller and smaller place, so as to get more and more worried about this particular item. It seems to be almost ridiculous that you can squeeze it to a numerical question that one thing is bigger than another. But there you are — it is bigger than any logical argument can produce, if you have this kind of logic.*"[4]

The maximally-entangled photon pair aside, there exist other entangled states that also demonstrate the incompatibility of quantum mechanics with LHVTs.[10-12] In Sec.3, we discuss the case of a photon pair in the Hardy state, and follow this in Sec.4 by examining the case of a triplet of photons in the Greenberger-Horne-Zeilinger (GHZ) thought experiment.

In photon-polarization-state studies, the Poincaré sphere serves as a powerful tool by mapping each and every state of polarization to a unique point on a unit-sphere's surface. Section 5 provides a detailed description of the Poincaré sphere in conjunction with the Stokes parameters, first defined in a basis of linearly-polarized states, then translated to a basis of right- and left-circularly-polarized states. This provides a segue into Sec.6, where the Bloch sphere, a close relative of the Poincaré sphere, is briefly introduced to help visualize the polarization states of spin-½ particles such as electrons and protons. The spin polarization of such particles can be examined with the aid of Stern-Gerlach analyzers,[1] which — much the same as polarizing beam-splitters used in photon experiments — can be oriented in three-dimensional space for purposes of separating spin-up and spin-down electrons (relative to an axis of the analyzer) by directing the particles of differing spins along divergent paths. For particles with nonzero mass (e.g., electrons, protons, neutrons, atoms), the mathematics of coordinate rotations is more nuanced than that needed for zero-mass particles (e.g., photons, neutrinos), which lack a rest frame. Thus, we devote Sec.7 to a systematic study of $2 \times 2$ unitary operators whose action on the spin state of a spin-½ particle at rest can bring about an arbitrary rotation of the coordinate system.[†]

In Sec.8, a pair of maximally entangled electrons is sent through two Stern-Gerlach analyzers with their axes oriented in different directions. For a particular setting of the analyzer angles, quantum mechanics predicts a probability of ¼ for the electrons emerging from the Stern-Gerlach devices to have anti-parallel spins. In contrast, LHVTs put the chances of such occurrences at greater than ⅓. Once again, the incompatibility of LHVTs with a description of Nature in quantum mechanical terms is brought into sharp focus.

Up to this point in the paper, we have managed to discuss the predictions of LHVTs on a case by case basis. Section 9 presents a formal (and more general) formulation of these theories, where one of Bell's inequalities (the one due to Clauser, Holt, Horne, and Shimony) is rigorously derived.

The unitary $2 \times 2$ rotation operator developed in Sec.7 for our subsequent study (in Sec.8) of a pair of entangled spin-½ particles, has numerous interesting properties and applications in quantum mechanics that go beyond the immediate concerns of the present paper. As an example, we show how the rotation operator helps to derive the spin angular momentum operator $\hat{\boldsymbol{S}} = \hat{S}_x\hat{\boldsymbol{x}} + \hat{S}_y\hat{\boldsymbol{y}} + \hat{S}_z\hat{\boldsymbol{z}}$ for spin-½ particles, then use the latter operator in Sec.10 to examine the singlet and triplet states of a pair of electrons.[1] This leads to a derivation of the $3 \times 3$ rotation operator for spin-1 particles of nonzero mass, thus providing a useful comparison with the $2 \times 2$ rotation operator for spin-1 photons mentioned toward the end of Sec.8.

[†] These $2 \times 2$ unitary operators act within the rest frame of two-state particles to bring about desired coordinate reorientations. However, they are not applicable to two-state particles with zero-mass, such as single photons, whose allowed rotations are restricted to those around the particle's propagation direction.[1]

With the $3 \times 3$ rotation operator for spin-1 particles at hand, we proceed to derive in Sec.11 the angular momentum operator $\hat{\boldsymbol{J}} = \hat{J}_x \hat{\boldsymbol{x}} + \hat{J}_y \hat{\boldsymbol{y}} + \hat{J}_z \hat{\boldsymbol{z}}$ for spin-1 particles of nonzero mass, then use the results to examine the states of a two-particle system consisting of a spin-1 particle $a$ and a spin-½ particle $b$. Once again, the methodology used and the insights gained from these last sections are in keeping with Feynman's unique approach to quantum mechanics.[1] The paper closes with a brief summary and a few concluding remarks in Sec.12.

**2. Maximally entangled photon pair**. The state of a maximally entangled pair of photons having linear polarization along the $x$ and $y$ axes is[4]

$$|\psi\rangle = \tfrac{1}{\sqrt{2}}(|x\rangle_1 |y\rangle_2 + |y\rangle_1 |x\rangle_2). \tag{1}$$

With reference to Fig.1, if analyzer 1 is rotated through an angle $\theta_1$ around the $z$-axis, photon 1 may be described as being in a superposition of $|x'\rangle_1$ and $|y'\rangle_1$, in which case the state of photon 2 automatically becomes a superposition of $|x\rangle_2$ and $|y\rangle_2$ corresponding to $|x'\rangle_2$ and $|y'\rangle_2$ associated with a rotated analyzer 2 through the *same* angle $\theta_1$, as follows:

$$\begin{aligned} |\psi\rangle &= \tfrac{1}{\sqrt{2}}[(\cos\theta_1 |x'\rangle_1 - \sin\theta_1 |y'\rangle_1)|y\rangle_2 + (\sin\theta_1 |x'\rangle_1 + \cos\theta_1 |y'\rangle_1)|x\rangle_2] \\ &= \tfrac{1}{\sqrt{2}}[|x'\rangle_1(\sin\theta_1 |x\rangle_2 + \cos\theta_1 |y\rangle_2) + |y'\rangle_1(\cos\theta_1 |x\rangle_2 - \sin\theta_1 |y\rangle_2)] \\ &= \tfrac{1}{\sqrt{2}}(|x'\rangle_1 |y'\rangle_2 + |y'\rangle_1 |x'\rangle_2). \end{aligned} \tag{2}$$

Clearly, when analyzers 1 and 2 are set to the *same* angle $\theta_1$, detectors 1 and 2 *both* detect a photon, or *both* fail to detect a photon, irrespective of the specific value of $\theta_1$.

---

**Digression**: For a single photon in the pure state $|\varphi\rangle = \alpha|x\rangle + \beta|y\rangle$, one can write $|\varphi\rangle = (\alpha\cos\theta + \beta\sin\theta)|x'\rangle - (\alpha\sin\theta - \beta\cos\theta)|y'\rangle$. Thus, when $\theta = \pm 90°$, $|\varphi\rangle = \pm(\beta|x'\rangle - \alpha|y'\rangle)$, indicating that a 90° rotation of the analyzer turns the probability $|\alpha|^2$ of passage into that of blockage, and vice-versa. Consistency requires that two successive 90° rotations restore the analyzer to its initial state. Indeed, when $\theta = \pm 180°$, $|\varphi\rangle = -(\alpha|x'\rangle + \beta|y'\rangle)$, which has the same probability $|\alpha|^2$ of passage and $|\beta|^2 = 1 - |\alpha|^2$ of blockage as the original setting at $\theta = 0°$.

---

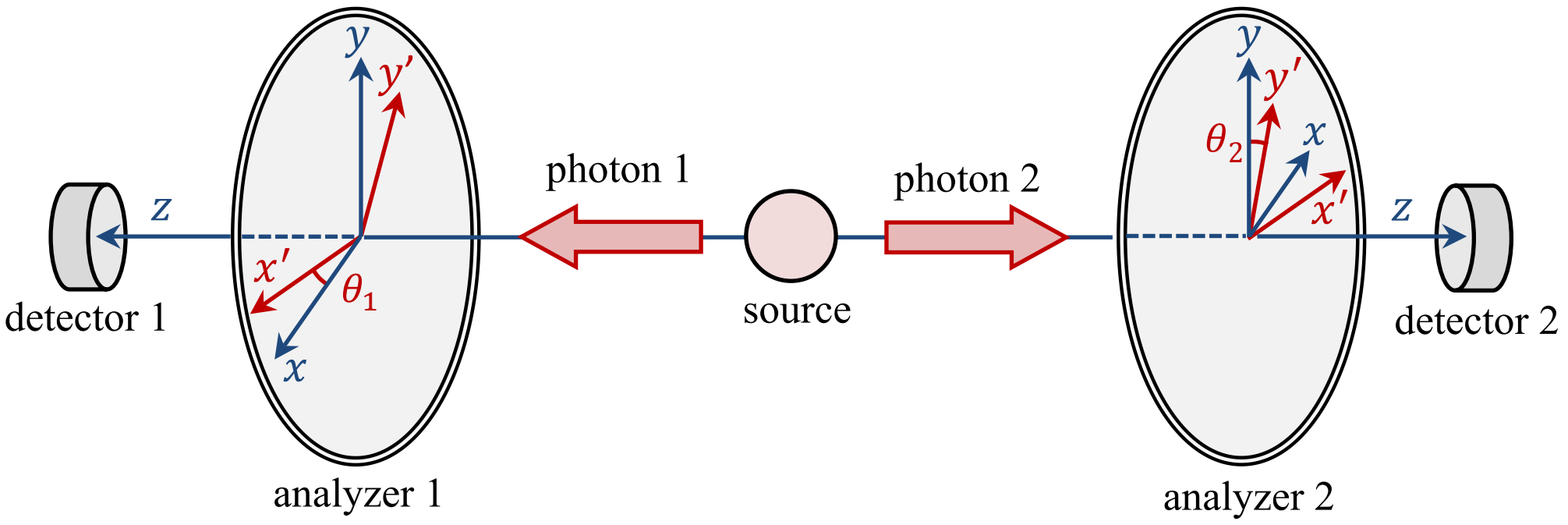

**Fig.1**. A source emits a pair of maximally entangled photons, one going to the left, the other to the right, toward detectors 1 and 2. (This could be due to a $3s \to 2p \to 1s$ transition in a hydrogen atom, resulting in two successive photoemissions.) Analyzers are placed before each detector, so that a photon polarized along the $x$-axis passes through to detector 1, whereas a photon polarized along $y$ is blocked. Similarly, a photon polarized along $y$ passes through to detector 2, while a photon polarized along $x$ is blocked. Each analyzer can be rotated independently of the other, so that when the transmission axis of the first analyzer is rotated through the angle $\theta_1$, that of the second analyzer is set to $\theta_2$. [The $xyz$ and $x'y'z'$ coordinate systems are right-handed (with the thumb along the $z$-axis). Both depicted angles $\theta_1$ and $\theta_2$ are taken to be positive.]

Next, assume analyzer 1 is set to $\theta_1$ while analyzer 2 is set to $\theta_2$. The entangled state $|\psi\rangle$ of the photon pair may now be written as

$$|\psi\rangle = \tfrac{1}{\sqrt{2}}[(\cos\theta_1\,|x'\rangle_1 - \sin\theta_1\,|y'\rangle_1)(-\sin\theta_2\,|x'\rangle_2 + \cos\theta_2\,|y'\rangle_2)$$
$$+(\sin\theta_1\,|x'\rangle_1 + \cos\theta_1\,|y'\rangle_1)(\cos\theta_2\,|x'\rangle_2 + \sin\theta_2\,|y'\rangle_2)]$$
$$= \tfrac{1}{\sqrt{2}}[\sin(\theta_1-\theta_2)\,|x'\rangle_1|x'\rangle_2 + \cos(\theta_1-\theta_2)\,|x'\rangle_1|y'\rangle_2$$
$$+\cos(\theta_1-\theta_2)\,|y'\rangle_1|x'\rangle_2 - \sin(\theta_1-\theta_2)\,|y'\rangle_1|y'\rangle_2]. \tag{3}$$

In this case, the probability that both detectors click, or that neither clicks, is $\cos^2(\theta_1-\theta_2)$, whereas the probability of only one detector clicking is $\sin^2(\theta_1-\theta_2)$.

---

**Digression**: The state $|\psi\rangle$ in Eq.(1) may be written in terms of the right and left circularly-polarized states (i.e., the "up" and "down" angular momentum states), where $|\uparrow\rangle = (|x\rangle + \mathrm{i}|y\rangle)/\sqrt{2}$ and $|\downarrow\rangle = (|x\rangle - \mathrm{i}|y\rangle)/\sqrt{2}$, as follows:

$$|\psi\rangle = \tfrac{1}{2\sqrt{2}}[(|\uparrow\rangle + |\downarrow\rangle)_1(|\uparrow\rangle - |\downarrow\rangle)_2 + (|\uparrow\rangle - |\downarrow\rangle)_1(|\uparrow\rangle + |\downarrow\rangle)_2] = \tfrac{1}{\sqrt{2}}(|\uparrow\rangle_1|\uparrow\rangle_2 - |\downarrow\rangle_1|\downarrow\rangle_2). \tag{4}$$

Note that the angular momentum of each photon in its up and down states is aligned with its own $\pm z$-axis, where $z$ is the direction of propagation. Thus, both the $|\uparrow\rangle_1|\uparrow\rangle_2$ and $|\downarrow\rangle_1|\downarrow\rangle_2$ states have zero net angular momentum. In spite of the distinguishability of photons of different color emitted in the $3s \to 2p \to 1s$ transition in a hydrogen atom, there is no *a priori* way to know whether the emitted pair is in the angular momentum state $|\uparrow\rangle_1|\uparrow\rangle_2$ or $|\downarrow\rangle_1|\downarrow\rangle_2$, hence the necessity of describing the state as a superposition of the two. For our pair of photons, not only is it necessary for the total $J_z$, whose operator is $\hat{J}_{z1} - \hat{J}_{z2}$, to vanish, but also the total angular momentum $\boldsymbol{J}$, associated with the operator $\hat{\boldsymbol{J}}\cdot\hat{\boldsymbol{J}} = \hat{\boldsymbol{J}}^2 = (\hat{J}_{z1} - \hat{J}_{z2})^2 = \hat{J}_{z1}^2 + \hat{J}_{z2}^2 - 2\hat{J}_{z1}\hat{J}_{z2}$, must vanish as well. This is the fundamental reason why the photon pair must be in the superposition state $|\psi\rangle$ given by Eq.(4) — the singlet state.

Given that photons are spin-1 particles, one might be curious as to why their spin angular momenta assume only the eigenvalues $J_z = \pm\hbar$; shouldn't there be a $J_z = 0$ eigenvalue as well? Here is how Feynman explains the absence of the corresponding quantum number $j = 0$: "But light is screwy; it has only two states. It does not have the zero case. This strange lack is related to the fact that light cannot stand still. For a particle of spin $j$ which is standing still, there must be the $2j+1$ possible states with values of $j_z$ going in steps of 1 from $-j$ to $+j$. But it turns out that for something of spin $j$ with zero mass only the states with the components $+j$ and $-j$ along the direction of motion exist. For example, light does not have three states, but only two — although a photon is still an object of spin one."[1]

---

**2.1. Density matrix**. The two-particle pure state $|\psi\rangle$ of Eq.(3) has a $4\times 4$ density matrix[13] (or density operator) consisting of 16 elements, as follows:

$$\hat{\rho}_{12}(\theta_1,\theta_2) = |\psi\rangle\langle\psi| = ½[\sin^2(\theta_1-\theta_2)\,|x'\rangle_1|x'\rangle_2\langle x'|_2\langle x'|_1$$
$$+\sin(\theta_1-\theta_2)\cos(\theta_1-\theta_2)\;|x'\rangle_1|x'\rangle_2\langle y'|_2\langle x'|_1$$
$$\vdots$$
$$-\sin(\theta_1-\theta_2)\cos(\theta_1-\theta_2)\,|y'\rangle_1|y'\rangle_2\langle x'|_2\langle y'|_1$$
$$+\sin^2(\theta_1-\theta_2)\,|y'\rangle_1|y'\rangle_2\langle y'|_2\langle y'|_1]. \tag{5}$$

The trace of $\hat{\rho}_{12}$ over particle 1 now yields the density matrix $\hat{\rho}_2$ of particle 2, namely,

$$\hat{\rho}_2(\theta_1,\theta_2) = \langle x'|_1|\psi\rangle\langle\psi||x'\rangle_1 + \langle y'|_1|\psi\rangle\langle\psi||y'\rangle_1 = ½|x'\rangle_2\langle x'|_2 + ½|y'\rangle_2\langle y'|_2. \tag{6}$$

This is the $2\times 2$ matrix $½\mathbb{I}$, showing that the state of particle 2 is maximally mixed, with $|x'\rangle$ and $|y'\rangle$ each having probability ½.

**2.2. A local hidden-variable theory**. Suppose each photon carries an instruction set as to which setting of an analyzer it is allowed to pass through, and which setting it is not.[4] Considering that the two (maximally entangled) photons behave in precisely the same way when the two analyzers are identically oriented (i.e., when $\theta_1 = \theta_2$), it is natural to assume that the instructions carried by the two photons are identical. However, in each photon-pair-emission event, the pair is expected to receive a new (perhaps randomly selected) instruction set.

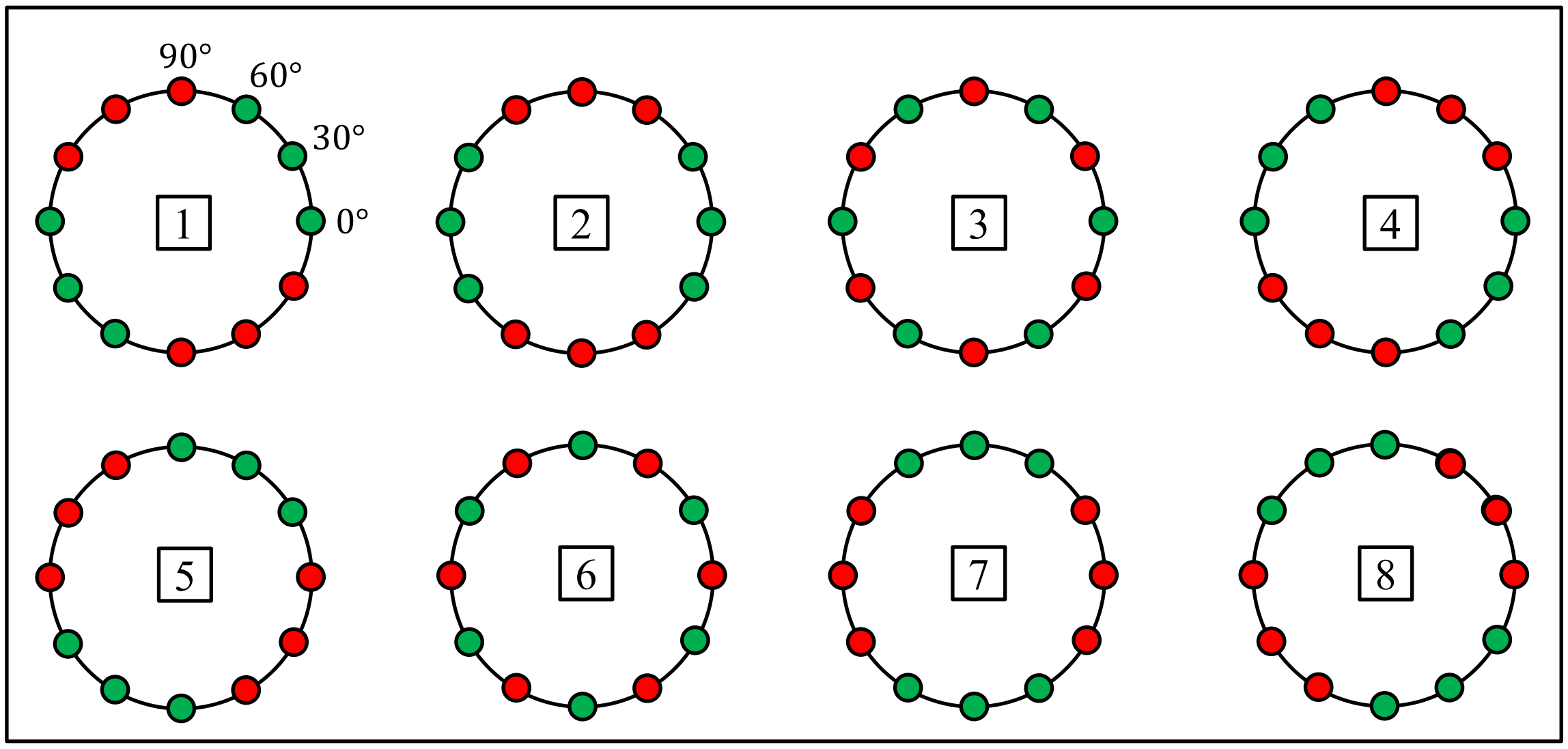

**Fig.2**. The two (maximally entangled) photons of Fig.1 are assumed to abide by identical instruction sets that specify the settings of their respective analyzer at which they are permitted to pass through to the corresponding detector. The instruction sets, although always identical for the two photons, change (perhaps randomly) from one emitted photon pair to the next. In each experiment, the angle $\theta_1$ of the first analyzer is set to a random multiple of 30° (relative to the $x$-axis), while the second analyzer is set to $\theta_2 = \theta_1 + 30°$ (relative to the $y$-axis). In each realization, the photons are instructed to pass or not pass through their respective analyzer when the analyzer is set at $\theta = 0°$, 30° and 60°. Considering that the rotation of an analyzer by 90° must result in the opposite outcome, the behavior of each photon for all the remaining angles (at integer-multiples of 30°) is automatically determined by their instructions for $\theta = 0°$, 30° and 60°. In the diagrams, a green dot at a given angle instructs the photon to pass through, whereas a red dot is an instruction to stop. There are numerous possible instruction sets, of course, but they can all be neatly divided into the depicted eight categories depending on the specific instructions they contain for the 0°, 30° and 60° analyzer settings. In the cases identified with the numerals 3 and 6, one detector always detects a photon and the other does not. For each of the remaining six cases, however, given that $\theta_1$ is randomly set to an integer-multiple of 30°, in eight out of twelve possible experiments (i.e., an average of two-thirds), the outputs of the two detectors will agree with each other. Consequently, no matter what fraction of the instruction sets falls into each of the eight possible categories, the fraction of experiments in which the readings of the two detectors agree with each other cannot exceed ⅔.

Now, in each repetition of our experiment, we randomly set the angle $\theta_1$ of the first analyzer to an integer-multiple of 30°, while fixing the second analyzer at $\theta_2 = \theta_1 + 30°$.[‡] Considering that the allowed orientation angles of both analyzers are integer-multiples of 30°, all conceivable instruction sets can be divided into the eight categories depicted in Fig.2. Here a green dot associated with a given angle ($\theta_1$ or $\theta_2$) indicates that the photon will pass through to the detector, whereas a red dot instructs the photon to stop. It is clear that, in cases marked as 3 and 6, one detector picks up a photon and the other does not, whereas in each of the remaining cases, for eight out of twelve possible settings of the analyzers, the outputs of the two detectors will agree with each other — that is, either both detectors pick up a photon or neither does. The probability of agreement between the two detector outputs (i.e., both receive a photon or neither does) is seen to be at most ⅔; that is, the probability of observing identical detector signals is below 67%. In contrast, the quantum mechanical theory of the preceding section predicts a coincidence rate of $\cos^2(\theta_1-\theta_2)=\cos^2(30°)=3/4$, whose validity has been amply confirmed by experiments.[14]. These results clearly indicate the incompatibility of quantum mechanics with the existence of local hidden variables.

[‡]To avoid the need to decide the values of $\theta_1$ and $\theta_2$ prior to the release of each photon pair, the two measurements can be done independently of each other, where, in each trial, the analyzer angles are set to random multiples of 30°. Afterward, when the experimenters come together, they compare only the results of those measurements that pertain to cases where $\theta_2 = \theta_1 + 30°$. Also, to ensure that coincidences are properly recorded, it is necessary in practice to replace the analyzers with polarizing beam-splitters, each followed by two photodetectors, one for detecting $x$-polarized photons, the other for $y$-polarized photons.

**Digression**: An alternative argument can be based on setting each analyzer (randomly and independently of the other) to $\theta=0°$, or 120°, or 240°. According to quantum mechanics, in cases where $|\theta_2-\theta_1|=120°$, the two detector signals must agree with a probability of $\cos^2(120°)=$ ¼. In contrast, should the photon pair emerge from the source with individual instruction sets (as described above), the chances of agreement between the detector signals must exceed 33%, in clear violation of the dictates of quantum mechanics.

**3. Photon pair in the Hardy state.**[12] The Hardy state of a hypothetical photon pair is given by

$$|\psi\rangle=\tfrac{1}{\sqrt{12}}\big(|x\rangle_1|x\rangle_2-|x\rangle_1|y\rangle_2-|y\rangle_1|x\rangle_2-3|y\rangle_1|y\rangle_2\big). \qquad (7)$$

In what follows, we shall assume that both analyzers transmit along their $x'$-axes and block along their $y'$-axes, with $x'y'$ rotated around $z$ through the angle $\theta$. (Viewed from the corresponding detector, each analyzer is rotated counterclockwise.) In the state $|\psi\rangle$ of Eq.(7), the probability of $|x\rangle_1|x\rangle_2$ is $1/12$, indicating that, when the analyzers are set to $\theta_1=\theta_2=0°$, there exists an 8.33% chance that detectors 1 and 2 will simultaneously pick up a photon. If analyzer 1 is rotated to $\theta_1=45°$, the state of the photon pair in terms of $|x'\rangle_1$ and $|y'\rangle_1$ becomes

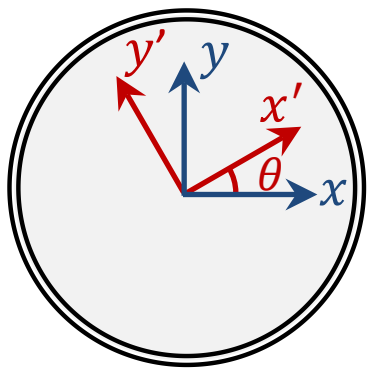

$$|\psi\rangle=\tfrac{1}{\sqrt{24}}\big[(|x'\rangle_1-|y'\rangle_1)|x\rangle_2-(|x'\rangle_1-|y'\rangle_1)|y\rangle_2-(|x'\rangle_1+|y'\rangle_1)|x\rangle_2$$
$$-3(|x'\rangle_1+|y'\rangle_1)|y\rangle_2\big]=-\tfrac{1}{\sqrt{6}}\big(2|x'\rangle_1|y\rangle_2+|y'\rangle_1|x\rangle_2+|y'\rangle_1|y\rangle_2\big). \qquad (8)$$

Considering that $|x'\rangle_1|x\rangle_2$ is missing from $|\psi\rangle$, there is no chance that the two detectors will simultaneously pick up photons when $\theta_1=45°$ and $\theta_2=0°$. Similarly, if analyzer 1 remains at $\theta_1=0°$ while analyzer 2 is rotated to $\theta_2=45°$, we will have

$$|\psi\rangle=\tfrac{1}{\sqrt{24}}\big[|x\rangle_1(|x'\rangle_2-|y'\rangle_2)-|x\rangle_1(|x'\rangle_2+|y'\rangle_2)-|y\rangle_1(|x'\rangle_2-|y'\rangle_2)$$
$$-3|y\rangle_1(|x'\rangle_2+|y'\rangle_2)\big]=-\tfrac{1}{\sqrt{6}}\big(|x\rangle_1|y'\rangle_2+2|y\rangle_1|x'\rangle_2+|y\rangle_1|y'\rangle_2\big). \qquad (9)$$

This time, $|x\rangle_1|x'\rangle_2$ drops out of $|\psi\rangle$, indicating the impossibility of simultaneous detection of both photons when $\theta_1=0°$ and $\theta_2=45°$. Finally, when the analyzers are set to $\theta_1=\theta_2=45°$, we will have

$$|\psi\rangle=\tfrac{1}{\sqrt{48}}\big[(|x'\rangle_1-|y'\rangle_1)(|x'\rangle_2-|y'\rangle_2)-(|x'\rangle_1-|y'\rangle_1)(|x'\rangle_2+|y'\rangle_2)$$
$$-(|x'\rangle_1+|y'\rangle_1)(|x'\rangle_2-|y'\rangle_2)-3(|x'\rangle_1+|y'\rangle_1)(|x'\rangle_2+|y'\rangle_2)\big]$$
$$=-\tfrac{1}{\sqrt{3}}\big(|x'\rangle_1|x'\rangle_2+|x'\rangle_1|y'\rangle_2+|y'\rangle_1|x'\rangle_2\big). \qquad (10)$$

In this case, $|y'\rangle_1|y'\rangle_2$ is missing from $|\psi\rangle$, indicating that at least one of the photons must be picked up by a detector when $\theta_1 = \theta_2 = 45°$.

Now, if we assume that each photon leaves the source with instructions as to how to behave upon reaching an analyzer, then all possible combinations of the instruction sets must adhere to the rules summarized in Table 1. In case $B$, corresponding to $(\theta_1, \theta_2) = (45°, 0°)$, combinations marked as $\times_B$ are prohibited; in case $C$, where $(\theta_1, \theta_2) = (0°, 45°)$, combinations marked as $\times_C$ are prohibited; and in case $D$, where $(\theta_1, \theta_2) = (45°, 45°)$, combination marked as $\times_D$ are prohibited. Under these circumstances, there is no chance of both detectors picking up photons in case $A$ corresponding to $(\theta_1, \theta_2) = (0°, 0°)$, hence the incompatibility of quantum mechanics with LHVTs.

| Photon 2, $\theta_2 = 0°/45°$ → <br> ↓ Photon 1, $\theta_1 = 0°/45°$ | pass/pass | stop/pass | pass/stop | stop/stop |
|---|---|---|---|---|
| pass/pass | $\times_{B,C}$ | $\times_C$ | $\times_B$ | |
| stop/pass | $\times_B$ | | $\times_B$ | |
| pass/stop | $\times_C$ | $\times_C$ | $\times_D$ | $\times_D$ |
| stop/stop | | | $\times_D$ | $\times_D$ |

**Table 1**. Certain combinations of the instruction sets carried by photons 1 and 2 are prohibited because simultaneous passage is disallowed in cases $B$ and $C$, while simultaneous blockage is disallowed in case $D$.

**4. Triplet of photons in the GHZ state.**[12] In the Greenberger-Horne-Zeilinger (GHZ) thought experiment, three photons are polarization-entangled in the following state:

$$|\psi\rangle = ½(|yyy\rangle_{1,2,3} - |yxx\rangle_{1,2,3} - |xyx\rangle_{1,2,3} - |xxy\rangle_{1,2,3}). \qquad (11)$$

The notation $|abc\rangle_{1,2,3}$ used here is an abbreviation for $|a\rangle_1|b\rangle_2|c\rangle_3$. Upon their release from a common source, the photons travel to three (mutually distant) detection stations, where each photon faces an analyzer followed by a detector. The analyzers placed in the paths of photons $1, 2, 3$ have their transmission axes rotated by $\theta_1, \theta_2, \theta_3$ from a common $x$-axis. Four different experiments are carried out with these three (entangled) photons, as follows:

***Case A***: $\theta_1 = \theta_2 = \theta_3 = 0°$. Considering the structure of $|\psi\rangle$, either none or exactly two photons are detected; i.e., the total number of photons detected in any such event is an even number.

***Case B***: $\theta_1 = \theta_2 = 45°$ while $\theta_3 = 0°$. The states of photons 1 and 2 must now be expressed in terms of the 45°-rotated axes $x'$ and $y'$, where $|x\rangle = \frac{1}{\sqrt{2}}(|x'\rangle - |y'\rangle)$ and $|y\rangle = \frac{1}{\sqrt{2}}(|x'\rangle + |y'\rangle)$. We will have

$$\begin{aligned}|\psi\rangle = ¼\big[&(|x'\rangle_1 + |y'\rangle_1)(|x'\rangle_2 + |y'\rangle_2)|y\rangle_3 - (|x'\rangle_1 + |y'\rangle_1)(|x'\rangle_2 - |y'\rangle_2)|x\rangle_3 \\ &-(|x'\rangle_1 - |y'\rangle_1)(|x'\rangle_2 + |y'\rangle_2)|x\rangle_3 - (|x'\rangle_1 - |y'\rangle_1)(|x'\rangle_2 - |y'\rangle_2)|y\rangle_3\big] \\ = ½(&|x'y'y\rangle_{1,2,3} + |y'x'y\rangle_{1,2,3} - |x'x'x\rangle_{1,2,3} + |y'y'x\rangle_{1,2,3}). \qquad (12)\end{aligned}$$

The total number of detected photons in this case will be either 1 or 3, i.e., an odd number.

***Case C***: $\theta_1 = \theta_3 = 45°$ while $\theta_2 = 0°$. Expressed in terms of $|x'\rangle_1$, $|y'\rangle_1$, $|x\rangle_2$, $|y\rangle_2$, $|x'\rangle_3$ and $|y'\rangle_3$, the state $|\psi\rangle$ now becomes

$$|\psi\rangle = ½(|x'yy'\rangle_{1,2,3} + |y'yx'\rangle_{1,2,3} - |x'xx'\rangle_{1,2,3} + |y'xy'\rangle_{1,2,3}). \quad (13)$$

As in the Case $B$, the total number of detected photons (either 1 or 3) will be an odd number.

***Case D***: $\theta_2 = \theta_3 = 45°$ while $\theta_1 = 0°$. Expressed in terms of $|x\rangle_1$, $|y\rangle_1$, $|x'\rangle_2$, $|y'\rangle_2$, $|x'\rangle_3$ and $|y'\rangle_3$ the state $|\psi\rangle$ becomes

$$|\psi\rangle = ½(|yx'y'\rangle_{1,2,3} + |yy'x'\rangle_{1,2,3} - |xx'x'\rangle_{1,2,3} + |xy'y'\rangle_{1,2,3}). \quad (14)$$

Again, the total number of detected photons (either 1 or 3) is seen to be an odd number.

An LHVT would equip each photon with instructions as to how to behave upon encountering an analyzer oriented at $\theta = 0°$ or 45°. Figure 3 shows 32 (out of a possible 64) combinations of such instruction sets in the form of triplets of colored cards, with the upper half of each card corresponding to $\theta = 0°$, and the lower half to $\theta = 45°$. The green color instructs the photon to pass through an analyzer set at $\theta$, whereas the red color tells it to stop. Considering that in the Case $A$ (where $\theta_1 = \theta_2 = \theta_3 = 0°$) an even number of photons must pass through to the detectors, the top-half colors of any 3-card set cannot be $GGG$, $GRR$, $RGR$, or $RRG$. This eliminates 32 out of 64 possible 3-card combinations, which explains their exclusion from Fig.3. None of the 3-card combinations that comply with the Case $A$ requirement (and, therefore, appear in Fig.3) complies with the expected behavior in the remaining cases $B$, $C$, and $D$. The symbol × placed below each 3-card set in Fig.3 is meant to cross-out that particular combination due to its violation of the requirements of cases $B$, $C$, or $D$, as indicated by the corresponding subscript of ×. Given that all possible 3-card combinations are thus rejected, there is no viable explanation that can be based on the existence of local hidden variables for the observed behavior of the entangled photons in the GHZ thought experiment.

**Fig.3** (next page). These 32 possible combinations of instruction sets carried by three photons reveal the impossibility of invoking the existence of local hidden variables to account for the actual behavior observed in the four experiments conducted with the photon triplet in the GHZ state $|\psi\rangle$. Each photon carries a card with its top and bottom halves colored in green (for pass) or red (for stop) when it encounters an analyzer set to $\theta = 0°$ or 45°. Considering that, in the state $|\psi\rangle$, either no photons or exactly two photons are allowed to pass through when the analyzers are set to $\theta_1 = \theta_2 = \theta_3 = 0°$ (Case $A$), the color combinations $GGG$, $GRR$, $RGR$, and $RRG$ for the top halves of a 3-card set need not be considered. This eliminates 32 out of 64 possible card combinations, which explains why none of the combinations depicted here violate the requirements for case $A$. In cases $B$, $C$, and $D$, where one analyzer is set to $\theta = 0°$ while the other two are at $\theta = 45°$, the GHZ state requires that either 1 or 3 photons pass through the analyzers. Each of the depicted 3-card combinations violates this requirement and is, therefore, crossed out with the symbol × placed below the combination. The subscript of × indicates the specific case ($B$, $C$, or $D$) which the combination violates.

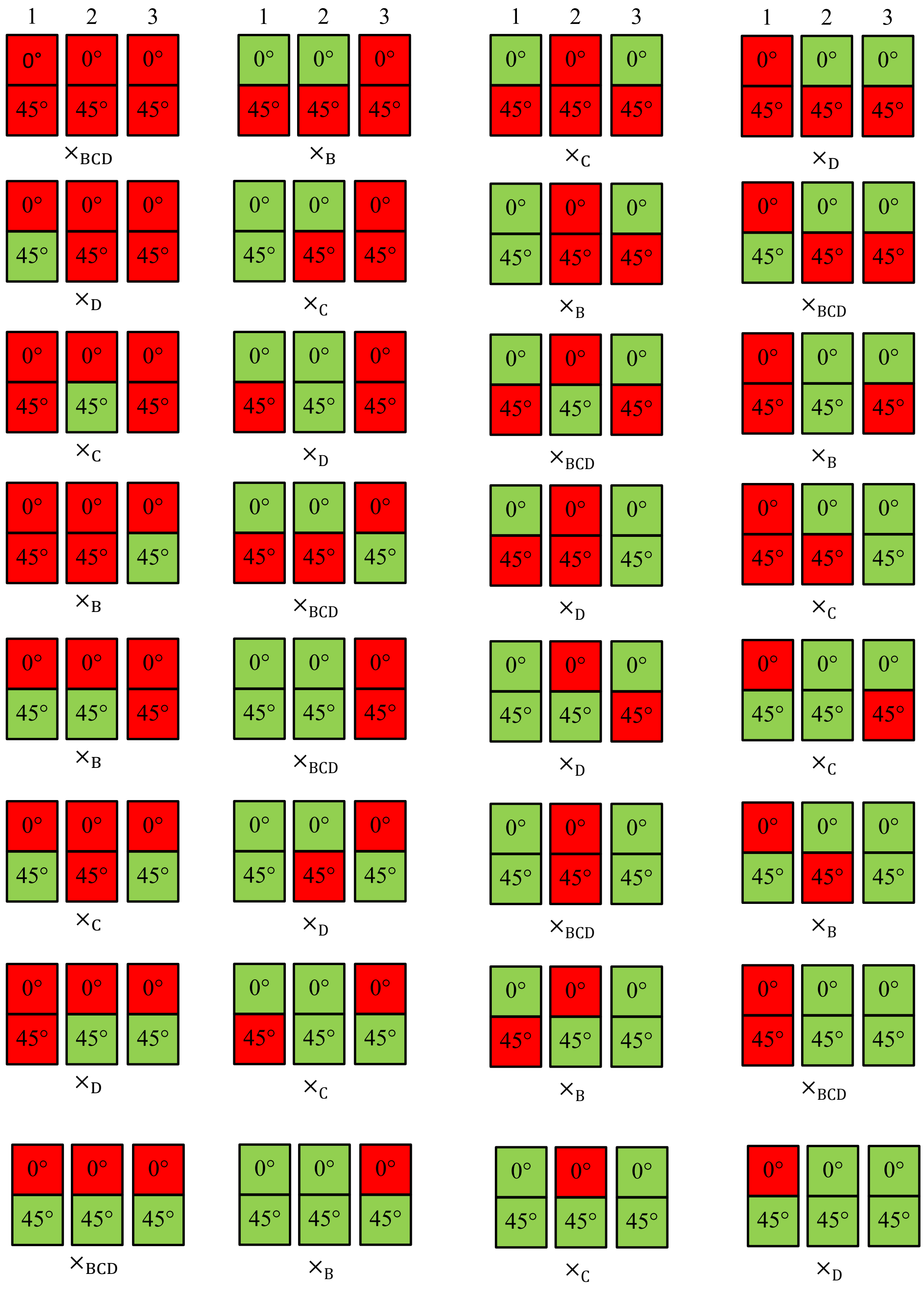

**5. The Poincaré sphere**. Let a photon be in a superposition of $|x\rangle$ and $|y\rangle$ polarization states with corresponding complex amplitudes $\alpha_x e^{\mathrm{i}\varphi_x}$ and $\alpha_y e^{\mathrm{i}\varphi_y}$, where $\alpha_x^2 + \alpha_y^2 = 1$. We write

$$|\psi\rangle = \alpha_x e^{\mathrm{i}\varphi_x}|x\rangle + \alpha_y e^{\mathrm{i}\varphi_y}|y\rangle. \tag{15}$$

The Stokes parameters for this state are defined as

$$S_0 = \alpha_x^2 + \alpha_y^2, \tag{16a}$$

$$S_1 = \alpha_x^2 - \alpha_y^2, \tag{16b}$$

$$S_2 = 2\alpha_x\alpha_y \cos(\varphi_y - \varphi_x), \tag{16c}$$

$$S_3 = 2\alpha_x\alpha_y \sin(\varphi_y - \varphi_x). \tag{16d}$$

Clearly, $S_1^2 + S_2^2 + S_3^2 = S_0^2 = 1$. The Poincaré sphere is a unit-radius sphere centered at $(x, y, z) = (0, 0, 0)$. As shown in Fig.4(a), the polarization state $|\psi\rangle$ is represented by a point on the sphere's surface located at $(x, y, z) = (S_1, S_2, S_3)$.

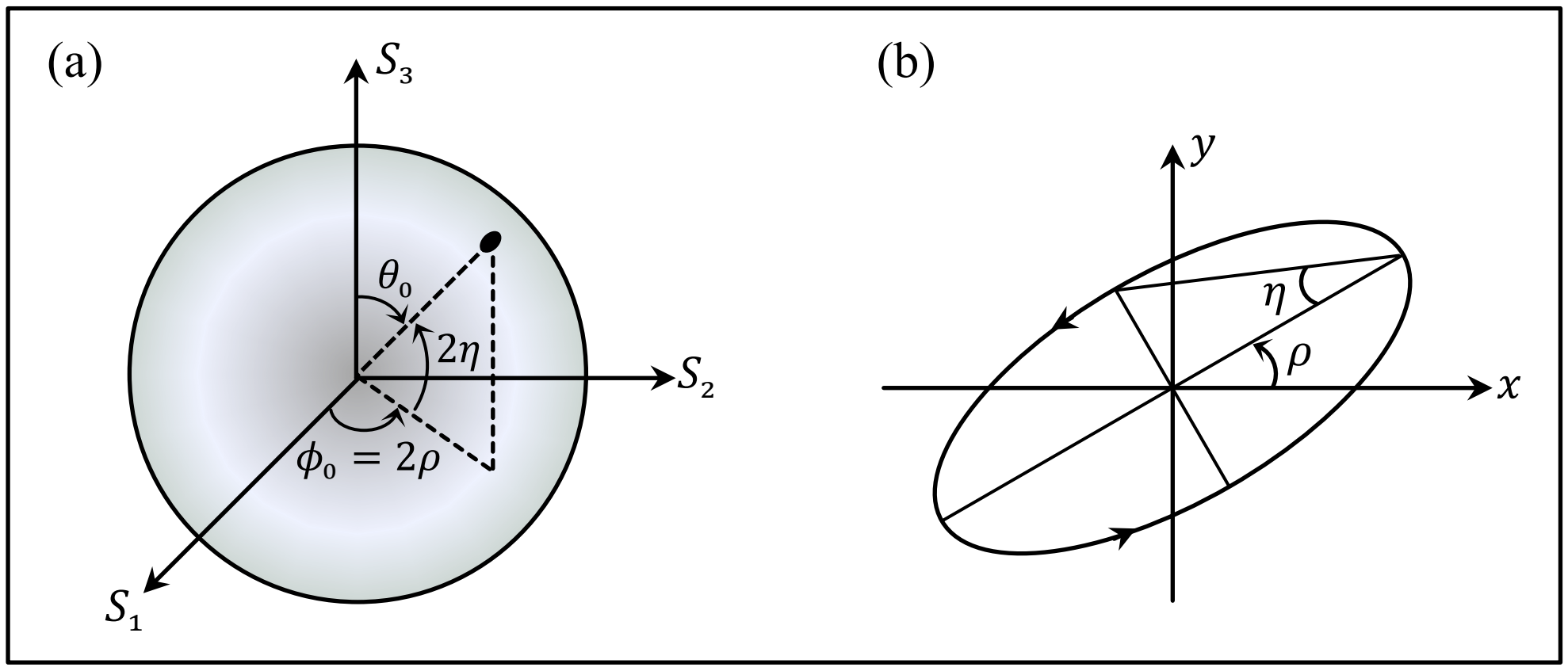

**Fig.4**. (a) The Poincaré sphere is a sphere of radius 1 centered at the origin of a coordinate system whose $S_1 S_2 S_3$ axes represent the Stokes parameters of a photon in the polarization state $|\psi\rangle$ of Eq.(15). The state is uniquely represented by a point on the surface of the Poincaré sphere. (b) The state $|\psi\rangle$ corresponds to an ellipse of polarization in the $xy$-plane, whose major axis makes an angle $\rho$ with the $x$-axis. The polarization ellipticity is specified by the angle $\eta$, which is the arctangent of the ratio of the minor to major axes of the ellipse. On the Poincaré sphere, the point representing $|\psi\rangle$ has longitude $2\rho$ and latitude $2\eta$. Alternatively, one may use the polar and azimuthal angles $(\theta_0, \phi_0)$ of a point on the sphere's surface to identify the corresponding state of polarization in a basis of right- and left-circular polarizations, namely, $|\text{RCP}\rangle = (|x\rangle + \mathrm{i}|y\rangle)/\sqrt{2}$ and $|\text{LCP}\rangle = (|x\rangle - \mathrm{i}|y\rangle)/\sqrt{2}$. Ignoring an insignificant phase factor, the polarization state can be written as $|\psi\rangle = \cos(½\theta_0)\,|\text{RCP}\rangle + \sin(½\theta_0)\, e^{\mathrm{i}\phi_0}|\text{LCP}\rangle$.

The ellipse of polarization is an ellipse in the $xy$-plane, having ellipticity $\eta$ and a major axis that is rotated through an angle $\rho$ from the $x$-axis toward the $y$-axis, as shown in Fig.4(b). The probability amplitudes along the major and minor axes of this ellipse are denoted by $(\cos\eta)e^{\mathrm{i}\chi}$ and $\mathrm{i}(\sin\eta)e^{\mathrm{i}\chi}$, respectively. Projection of the above amplitudes onto $|x\rangle$ and $|y\rangle$ yields

$$\cos\rho\cos\eta\, e^{\mathrm{i}\chi} - \mathrm{i}\sin\rho\sin\eta\, e^{\mathrm{i}\chi} = \alpha_x e^{\mathrm{i}\varphi_x}, \tag{17a}$$

$$\sin\rho\cos\eta\, e^{\mathrm{i}\chi} + \mathrm{i}\cos\rho\sin\eta\, e^{\mathrm{i}\chi} = \alpha_y e^{\mathrm{i}\varphi_y}. \tag{17b}$$

From Eqs.(16b), (17a), and (17b), we find

$$S_1 = \alpha_x^2 - \alpha_y^2 = |\cos\rho\cos\eta - \mathrm{i}\sin\rho\sin\eta|^2 - |\sin\rho\cos\eta + \mathrm{i}\cos\rho\sin\eta|^2 = \cos(2\eta)\cos(2\rho). \tag{18}$$

Also, upon multiplying the conjugate of Eq.(17a) into Eq.(17b), we arrive at

$$(\cos\rho\cos\eta + \mathrm{i}\sin\rho\sin\eta)(\sin\rho\cos\eta + \mathrm{i}\cos\rho\sin\eta) = \alpha_x\alpha_y e^{\mathrm{i}(\varphi_y - \varphi_x)}. \tag{19}$$

Equating the real and imaginary parts on both sides of Eq.(19) now yields

$$S_2 = \cos(2\eta)\sin(2\rho), \tag{20}$$

$$S_3 = \sin(2\eta). \tag{21}$$

Equations (18), (20), (21) reveal the longitude and latitude of the point $\boldsymbol{S} = S_1\hat{\boldsymbol{x}} + S_2\hat{\boldsymbol{y}} + S_3\hat{\boldsymbol{z}}$ on the surface of the Poincaré sphere to be $2\rho$ and $2\eta$, respectively.

---

**Digression**: Suppose an analyzer is placed in the path of the photon, perpendicular to the $z$-axis and oriented with its transmission axis at an angle $\theta$ with the $x$-axis. In the equatorial plane of the Poincaré sphere, the unit-vector $\boldsymbol{A} = \cos(2\theta)\,\hat{\boldsymbol{x}} + \sin(2\theta)\,\hat{\boldsymbol{y}}$ represents the transmission axis. The probability of transmission through the analyzer is readily seen to be

$$\begin{aligned} P_{\text{trans}} &= |\alpha_x e^{\mathrm{i}\varphi_x}\cos\theta + \alpha_y e^{\mathrm{i}\varphi_y}\sin\theta|^2 \\ &= \alpha_x^2\cos^2\theta + \alpha_y^2\sin^2\theta + 2\alpha_x\alpha_y\cos(\varphi_x - \varphi_y)\sin\theta\cos\theta \\ &= ½(\alpha_x^2 + \alpha_y^2) + ½(\alpha_x^2 - \alpha_y^2)\cos(2\theta) + \alpha_x\alpha_y\cos(\varphi_x - \varphi_y)\sin(2\theta) \\ &= ½[1 + S_1\cos(2\theta) + S_2\sin(2\theta)] = ½(1 + \boldsymbol{S}\cdot\boldsymbol{A}). \end{aligned} \tag{22}$$

---

If the photon's state is expressed as a superposition of right- and left-circular polarization states, we will have

$$\alpha_x e^{\mathrm{i}\varphi_x}|x\rangle + \alpha_y e^{\mathrm{i}\varphi_y}|y\rangle = ½(\alpha_x e^{\mathrm{i}\varphi_x} - \mathrm{i}\alpha_y e^{\mathrm{i}\varphi_y})(|x\rangle + \mathrm{i}|y\rangle) + ½(\alpha_x e^{\mathrm{i}\varphi_x} + \mathrm{i}\alpha_y e^{\mathrm{i}\varphi_y})(|x\rangle - \mathrm{i}|y\rangle), \tag{23}$$

where the amplitudes of the right- and left-circular polarization states, namely, $(|x\rangle \pm \mathrm{i}|y\rangle)/\sqrt{2}$, are

$$\beta_{\text{RCP}} = \frac{1}{\sqrt{2}}(\alpha_x e^{\mathrm{i}\varphi_x} - \mathrm{i}\alpha_y e^{\mathrm{i}\varphi_y}), \tag{24a}$$

$$\beta_{\text{LCP}} = \frac{1}{\sqrt{2}}(\alpha_x e^{\mathrm{i}\varphi_x} + \mathrm{i}\alpha_y e^{\mathrm{i}\varphi_y}). \tag{24b}$$

The Stokes parameters can now be related to the circular polarization amplitudes, as follows:

$$2\beta_{\text{LCP}}\beta_{\text{RCP}}^* = \alpha_x^2 - \alpha_y^2 + \mathrm{i}2\alpha_x\alpha_y\cos(\varphi_y - \varphi_x) = S_1 + \mathrm{i}S_2, \tag{25a}$$

$$|\beta_{\text{RCP}}|^2 - |\beta_{\text{LCP}}|^2 = 2\alpha_x\alpha_y\sin(\varphi_y - \varphi_x) = S_3. \tag{25b}$$

Apparently, the same Stokes parameters, albeit rearranged, are obtained when the state of the photon is described in a different orthonormal basis. Defining the angles $\theta_0$ and $\phi_0 = \varphi_{\text{LCP}} - \varphi_{\text{RCP}}$ such that $\beta_{\text{RCP}} = \cos(½\theta_0)\exp(\mathrm{i}\varphi_{\text{RCP}})$ and $\beta_{\text{LCP}} = \sin(½\theta_0)\exp(\mathrm{i}\varphi_{\text{LCP}})$, it is seen that $S_1 = \sin\theta_0\cos\phi_0$, $S_2 = \sin\theta_0\sin\phi_0$, and $S_3 = \cos\theta_0$. The photon's state is thus associated with the point $(\theta_0, \phi_0)$ on the surface of the Poincaré sphere, where, aside from an overall phase factor, $|\psi\rangle = \cos(½\theta_0)\,|\text{RCP}\rangle + \sin(½\theta_0)\,e^{\mathrm{i}\phi_0}|\text{LCP}\rangle$. Referring to Fig.4(a), the state at the sphere's north pole, where $\theta_0 = 0°$, is RCP, whereas that at the south pole, corresponding to $\theta_0 = 180°$, is LCP.

**6. Spin-1 photons, spin-½ electrons, and the Bloch sphere**. Writing the states of right and left circularly-polarized photons as $|\pm\rangle = \frac{1}{\sqrt{2}}(|x\rangle \pm \mathrm{i}|y\rangle)$, where $|\mathrm{RCP}\rangle = |+\rangle = |\uparrow\rangle$ and $|\mathrm{LCP}\rangle = |-\rangle = |\downarrow\rangle$, we note that the Stokes parameters for the $|\pm\rangle$ states are $S = (S_1, S_2, S_3) = (0, 0, \pm 1)$. These spin angular momentum eigenstates, each having their angular momentum aligned with or against the $z$-axis, appear at the north and south poles of the Poincaré sphere. Let $|\psi\rangle = \beta_+|+\rangle + \beta_-|-\rangle$. Defining $\beta_+ = \cos(\vartheta)\exp(\mathrm{i}\varphi_+)$ and $\beta_- = \sin(\vartheta)\exp(\mathrm{i}\varphi_-)$, we may write

$$|x\rangle = \tfrac{1}{\sqrt{2}}(|+\rangle + |-\rangle) \qquad \rightarrow \qquad \vartheta = 45°, \varphi_+ = 0°, \varphi_- = 0°. \tag{26}$$

$$|y\rangle = -\tfrac{\mathrm{i}}{\sqrt{2}}(|+\rangle - |-\rangle) \qquad \rightarrow \qquad \vartheta = 45°, \varphi_+ = -90°, \varphi_- = 90°. \tag{27}$$

This suggests that the conventional spherical coordinates of the various points on the Poincaré sphere are $(\theta_0, \phi_0) = (2\vartheta, \varphi_- - \varphi_+)$. Writing the general polarization state of a photon as $|\psi\rangle = \exp(\mathrm{i}\varphi_+)\left[\cos(½\theta_0)\,|+\rangle + \sin(½\theta_0)\,e^{\mathrm{i}\phi_0}|-\rangle\right]$ and proceeding to ignore the overall phase factor, it is seen that, at the north pole, where $\theta_0 = 0°$, $|\psi\rangle = |+\rangle$, whereas at the south pole, where $\theta_0 = 180°$, $|\psi\rangle = |-\rangle$. On the positive $x$-axis, $(\theta_0, \phi_0) = (90°, 0°)$ and $|\psi\rangle = |x\rangle$, while on the negative $x$-axis (corresponding to $\rho = 90°, \eta = 0°$), where $(\theta_0, \phi_0) = (90°, 180°)$, $|\psi\rangle = |y\rangle$.

For spin-½ particles such as electrons and protons, it is customary to use the Bloch sphere instead of Poincaré's sphere. The Bloch sphere is a sphere of radius 1, centered at the origin of an $xyz$ coordinate system. The general state of a spin-½ particle expressed relative to $\hat{\boldsymbol{z}}$ is written

$$|\psi\rangle = \alpha_1|\uparrow\rangle + \alpha_2|\downarrow\rangle = e^{\mathrm{i}\chi_0}\left[\cos(½\theta_0)\,|\uparrow\rangle + \sin(½\theta_0)\,e^{\mathrm{i}\phi_0}|\downarrow\rangle\right]. \tag{28}$$

Ignoring the absolute phase, the state along $\hat{\boldsymbol{x}}$, where $(\theta_0, \phi_0) = (90°, 0°)$, becomes $\frac{1}{\sqrt{2}}(|\uparrow\rangle + |\downarrow\rangle)$. Similarly, the state along $\hat{\boldsymbol{y}}$, where $(\theta_0, \phi_0) = (90°, 90°)$, becomes $\frac{1}{\sqrt{2}}(|\uparrow\rangle + \mathrm{i}|\downarrow\rangle)$.

**7. Unitary transformations**. To discuss the effect of coordinate rotations on a spin-½ particle with nonzero mass (e.g., an electron or a proton), one must resort to $2 \times 2$ unitary transformations whose general structure and properties are the subject of the present section.[12] We begin by introducing the $2 \times 2$ identity matrix $\hat{\mathbb{I}}$ and the Pauli matrices $\hat{\sigma}_x, \hat{\sigma}_y, \hat{\sigma}_z$, which have the following properties:

$$\hat{\mathbb{I}} = \begin{pmatrix} 1 & 0 \\ 0 & 1 \end{pmatrix}, \qquad \hat{\sigma}_x = \begin{pmatrix} 0 & 1 \\ 1 & 0 \end{pmatrix}, \qquad \hat{\sigma}_y = \begin{pmatrix} 0 & -\mathrm{i} \\ \mathrm{i} & 0 \end{pmatrix}, \qquad \hat{\sigma}_z = \begin{pmatrix} 1 & 0 \\ 0 & -1 \end{pmatrix}. \tag{29a}$$

$$\hat{\sigma}_x\hat{\sigma}_y = -\hat{\sigma}_y\hat{\sigma}_x = \mathrm{i}\hat{\sigma}_z, \qquad \hat{\sigma}_y\hat{\sigma}_z = -\hat{\sigma}_z\hat{\sigma}_y = \mathrm{i}\hat{\sigma}_x, \qquad \hat{\sigma}_z\hat{\sigma}_x = -\hat{\sigma}_x\hat{\sigma}_z = \mathrm{i}\hat{\sigma}_y. \tag{29b}$$

$$\hat{\sigma}_x^2 = \hat{\sigma}_y^2 = \hat{\sigma}_z^2 = \hat{\mathbb{I}}. \tag{29c}$$

Let $\vec{\boldsymbol{a}} = a_x\hat{\boldsymbol{x}} + a_y\hat{\boldsymbol{y}} + a_z\hat{\boldsymbol{z}}$ and $\vec{\boldsymbol{b}} = b_x\hat{\boldsymbol{x}} + b_y\hat{\boldsymbol{y}} + b_z\hat{\boldsymbol{z}}$ be two arbitrary vectors in three-dimensional Euclidean space. Using the identities in Eqs.(29b) and (29c) it is easy to show that

$$(\vec{\boldsymbol{a}} \cdot \hat{\boldsymbol{\sigma}})(\vec{\boldsymbol{b}} \cdot \hat{\boldsymbol{\sigma}}) = (\vec{\boldsymbol{a}} \cdot \vec{\boldsymbol{b}})\hat{\mathbb{I}} + \mathrm{i}(\vec{\boldsymbol{a}} \times \vec{\boldsymbol{b}}) \cdot \hat{\boldsymbol{\sigma}}. \tag{30}$$

(This identity holds even if the components of $\vec{\boldsymbol{a}}$ and $\vec{\boldsymbol{b}}$ are complex, although, in what follows, $\vec{\boldsymbol{a}}$ and $\vec{\boldsymbol{b}}$ are assumed to be real vectors.) Now, with the aid of four arbitrary complex numbers $u_0, u_1, u_2, u_3$, any $2 \times 2$ matrix (or operator) can be written as a linear combination of $\hat{\sigma}_x, \hat{\sigma}_y, \hat{\sigma}_z$, and $\hat{\mathbb{I}}$; that is,

$$\hat{u} = u_0\hat{\mathbb{I}} + u_1\hat{\sigma}_x + u_2\hat{\sigma}_y + u_3\hat{\sigma}_z = \begin{pmatrix} u_0 + u_3 & u_1 - \mathrm{i}u_2 \\ u_1 + \mathrm{i}u_2 & u_0 - u_3 \end{pmatrix} = \begin{pmatrix} u_{11} & u_{12} \\ u_{21} & u_{22} \end{pmatrix}. \tag{31}$$

Our goal is to specify $u_0, u_1, u_2, u_3$ such that $\hat{u}$ is a general-purpose unitary operator; that is, $\hat{u}\hat{u}^\dagger = \hat{u}^\dagger\hat{u} = \hat{\mathbb{I}}$. We start by restricting $u_0$ to real values, which renders the resulting $\hat{u}$ a general-purpose unitary operator to within an overall phase-factor $\exp(\mathrm{i}\varphi_0)$. Defining the complex vector $\vec{\boldsymbol{u}} = \vec{\boldsymbol{u}_r} + \mathrm{i}\vec{\boldsymbol{u}_\imath} = u_1\hat{\boldsymbol{x}} + u_2\hat{\boldsymbol{y}} + u_3\hat{\boldsymbol{z}}$ and the (Hermitian) vector operator $\hat{\boldsymbol{\sigma}} = \hat{\sigma}_x\hat{\boldsymbol{x}} + \hat{\sigma}_y\hat{\boldsymbol{y}} + \hat{\sigma}_z\hat{\boldsymbol{z}}$, we write

$$\begin{aligned}
\hat{u}\hat{u}^\dagger &= \left[u_0\hat{\mathbb{I}} + (\vec{\boldsymbol{u}_r} + \mathrm{i}\vec{\boldsymbol{u}_\imath})\cdot\hat{\boldsymbol{\sigma}}\right]\left[u_0\hat{\mathbb{I}} + (\vec{\boldsymbol{u}_r} - \mathrm{i}\vec{\boldsymbol{u}_\imath})\cdot\hat{\boldsymbol{\sigma}}\right] \\
&= u_0^2\hat{\mathbb{I}} + 2u_0(\vec{\boldsymbol{u}_r}\cdot\hat{\boldsymbol{\sigma}}) + (\vec{\boldsymbol{u}_r}\cdot\hat{\boldsymbol{\sigma}})(\vec{\boldsymbol{u}_r}\cdot\hat{\boldsymbol{\sigma}}) - \mathrm{i}(\vec{\boldsymbol{u}_r}\cdot\hat{\boldsymbol{\sigma}})(\vec{\boldsymbol{u}_\imath}\cdot\hat{\boldsymbol{\sigma}}) + \mathrm{i}(\vec{\boldsymbol{u}_\imath}\cdot\hat{\boldsymbol{\sigma}})(\vec{\boldsymbol{u}_r}\cdot\hat{\boldsymbol{\sigma}}) \\
&\quad + (\vec{\boldsymbol{u}_\imath}\cdot\hat{\boldsymbol{\sigma}})(\vec{\boldsymbol{u}_\imath}\cdot\hat{\boldsymbol{\sigma}}) \\
&= (u_0^2 + \vec{\boldsymbol{u}_r}\cdot\vec{\boldsymbol{u}_r} + \vec{\boldsymbol{u}_\imath}\cdot\vec{\boldsymbol{u}_\imath})\hat{\mathbb{I}} + 2u_0(\vec{\boldsymbol{u}_r}\cdot\hat{\boldsymbol{\sigma}}) + 2(\vec{\boldsymbol{u}_r}\times\vec{\boldsymbol{u}_\imath})\cdot\hat{\boldsymbol{\sigma}}.
\end{aligned} \tag{32a}$$

$$\hat{u}^\dagger\hat{u} = (u_0^2 + \vec{\boldsymbol{u}_r}\cdot\vec{\boldsymbol{u}_r} + \vec{\boldsymbol{u}_\imath}\cdot\vec{\boldsymbol{u}_\imath})\hat{\mathbb{I}} + 2u_0(\vec{\boldsymbol{u}_r}\cdot\hat{\boldsymbol{\sigma}}) - 2(\vec{\boldsymbol{u}_r}\times\vec{\boldsymbol{u}_\imath})\cdot\hat{\boldsymbol{\sigma}}. \tag{32b}$$

Since unitarity demands that $\hat{u}\hat{u}^\dagger = \hat{u}^\dagger\hat{u} = \hat{\mathbb{I}}$, we must have $\vec{\boldsymbol{u}_r}\times\vec{\boldsymbol{u}_\imath} = 0$; that is, $\vec{\boldsymbol{u}_r}$ and $\vec{\boldsymbol{u}_\imath}$ are required to be parallel vectors. Now, if $u_0 \neq 0$, we must set $\vec{\boldsymbol{u}_r} = 0$ to eliminate the term $2u_0(\vec{\boldsymbol{u}_r}\cdot\hat{\boldsymbol{\sigma}})$ from Eq.(32) — otherwise, this term would prevent $\hat{u}\hat{u}^\dagger$ and $\hat{u}^\dagger\hat{u}$ from equaling the identity matrix. If $u_0$ happens to be zero, we may retain $\vec{\boldsymbol{u}_r}$, but, recalling that $\vec{\boldsymbol{u}_r} \parallel \vec{\boldsymbol{u}_\imath}$, the only effect of $\vec{\boldsymbol{u}_r}$ in this case would be to multiply $\hat{u} = (\vec{\boldsymbol{u}_r} + \mathrm{i}\vec{\boldsymbol{u}_\imath})\cdot\hat{\boldsymbol{\sigma}}$ with a phase-factor, which we have been ignoring anyway. So, once again, $\vec{\boldsymbol{u}_r} = 0$ will be an acceptable choice. We thus set $\hat{u} = u_0\hat{\mathbb{I}} + \mathrm{i}\vec{\boldsymbol{u}_\imath}\cdot\hat{\boldsymbol{\sigma}}$ and require that $u_0^2 + \vec{\boldsymbol{u}_\imath}\cdot\vec{\boldsymbol{u}_\imath} = 1$ to guarantee that $\hat{u}\hat{u}^\dagger = \hat{u}^\dagger\hat{u} = \hat{\mathbb{I}}$ is satisfied.

Next, setting $u_0 = \cos(\theta/2)$ and $\vec{\boldsymbol{u}_\imath} = \sin(\theta/2)\vec{\boldsymbol{n}}$, where $\theta$ is an arbitrary angle and $\vec{\boldsymbol{n}} = n_x\hat{\boldsymbol{x}} + n_y\hat{\boldsymbol{y}} + n_z\hat{\boldsymbol{z}}$ is an arbitrary real-valued unit-vector in three-dimensional space, we arrive at the following expression for a $2\times 2$ unitary operator that, aside from a constant phase-factor, is completely general:

$$\hat{u} = \cos(\theta/2)\,\hat{\mathbb{I}} + \mathrm{i}\sin(\theta/2)\,\vec{\boldsymbol{n}}\cdot\hat{\boldsymbol{\sigma}}. \tag{33}$$

---

**Digression**: Occasionally, it is convenient to express the above unitary operator as $\hat{u} = \exp[½\mathrm{i}\theta(\vec{\boldsymbol{n}}\cdot\hat{\boldsymbol{\sigma}})]$, which is easily proven via the following identities:

$$\exp(A) = \textstyle\sum_{k=0}^{\infty} A^k/k!, \tag{34a}$$

$$(\vec{\boldsymbol{n}}\cdot\hat{\boldsymbol{\sigma}})^2 = (\vec{\boldsymbol{n}}\cdot\vec{\boldsymbol{n}})\hat{\mathbb{I}} + \mathrm{i}(\vec{\boldsymbol{n}}\times\vec{\boldsymbol{n}})\cdot\hat{\boldsymbol{\sigma}} = \hat{\mathbb{I}}, \tag{34b}$$

$$\sin x = x\textstyle\sum_{k=0}^{\infty}(-x^2)^k/(2k+1)!, \tag{34c}$$

$$\cos x = \textstyle\sum_{k=0}^{\infty}(-x^2)^k/(2k)!. \tag{34d}$$

---

Let us now consider the effect of our unitary transformation $\hat{u}$ on the operator $\vec{\boldsymbol{a}}\cdot\hat{\boldsymbol{\sigma}}$, which is defined via the real-valued but otherwise arbitrary vector $\vec{\boldsymbol{a}} = a_x\hat{\boldsymbol{x}} + a_y\hat{\boldsymbol{y}} + a_z\hat{\boldsymbol{z}}$. We find

$$\begin{aligned}
\hat{u}(\vec{\boldsymbol{a}}\cdot\hat{\boldsymbol{\sigma}})\hat{u}^\dagger &= \left[\cos(\theta/2)\,\hat{\mathbb{I}} + \mathrm{i}\sin(\theta/2)\,\vec{\boldsymbol{n}}\cdot\hat{\boldsymbol{\sigma}}\right](\vec{\boldsymbol{a}}\cdot\hat{\boldsymbol{\sigma}})\left[\cos(\theta/2)\,\hat{\mathbb{I}} - \mathrm{i}\sin(\theta/2)\,\vec{\boldsymbol{n}}\cdot\hat{\boldsymbol{\sigma}}\right] \\
&= \cos^2(\theta/2)\,(\vec{\boldsymbol{a}}\cdot\hat{\boldsymbol{\sigma}}) - \mathrm{i}\cos(\theta/2)\sin(\theta/2)\,(\vec{\boldsymbol{a}}\cdot\hat{\boldsymbol{\sigma}})(\vec{\boldsymbol{n}}\cdot\hat{\boldsymbol{\sigma}}) \\
&\quad + \mathrm{i}\sin(\theta/2)\cos(\theta/2)\,(\vec{\boldsymbol{n}}\cdot\hat{\boldsymbol{\sigma}})(\vec{\boldsymbol{a}}\cdot\hat{\boldsymbol{\sigma}}) + \sin^2(\theta/2)\,(\vec{\boldsymbol{n}}\cdot\hat{\boldsymbol{\sigma}})\underbrace{(\vec{\boldsymbol{a}}\cdot\hat{\boldsymbol{\sigma}})(\vec{\boldsymbol{n}}\cdot\hat{\boldsymbol{\sigma}})}_{\text{use Eq.(30)}}
\end{aligned}$$

$$= \cos^2(\theta/2)\,\vec{\boldsymbol{a}}\cdot\hat{\boldsymbol{\sigma}} + \sin\theta\,(\vec{\boldsymbol{a}}\times\vec{\boldsymbol{n}})\cdot\hat{\boldsymbol{\sigma}}$$

$$+\sin^2(\theta/2)\,(\vec{\boldsymbol{n}}\cdot\hat{\boldsymbol{\sigma}})\big[(\vec{\boldsymbol{a}}\cdot\vec{\boldsymbol{n}})\hat{\mathbb{I}} + \mathrm{i}(\vec{\boldsymbol{a}}\times\vec{\boldsymbol{n}})\cdot\hat{\boldsymbol{\sigma}}\big]$$

$$= \cos^2(\theta/2)\,\vec{\boldsymbol{a}}\cdot\hat{\boldsymbol{\sigma}} + \sin\theta\,(\vec{\boldsymbol{a}}\times\vec{\boldsymbol{n}})\cdot\hat{\boldsymbol{\sigma}} + \sin^2(\theta/2)\,(\vec{\boldsymbol{a}}\cdot\vec{\boldsymbol{n}})(\vec{\boldsymbol{n}}\cdot\hat{\boldsymbol{\sigma}})$$

$$+\mathrm{i}\sin^2(\theta/2)\,(\vec{\boldsymbol{n}}\cdot\hat{\boldsymbol{\sigma}})[(\vec{\boldsymbol{a}}\times\vec{\boldsymbol{n}})\cdot\hat{\boldsymbol{\sigma}}]$$ ← use Eq.(30)

$$= [\cos^2(\theta/2)\,\vec{\boldsymbol{a}} + \sin\theta\,(\vec{\boldsymbol{a}}\times\vec{\boldsymbol{n}}) + \sin^2(\theta/2)\,(\vec{\boldsymbol{a}}\cdot\vec{\boldsymbol{n}})\vec{\boldsymbol{n}}]\cdot\hat{\boldsymbol{\sigma}}$$

$$+\mathrm{i}\sin^2(\theta/2)\,\{[\underbrace{\vec{\boldsymbol{n}}\cdot(\vec{\boldsymbol{a}}\times\vec{\boldsymbol{n}})}_{0}]\hat{\mathbb{I}} + \mathrm{i}[\vec{\boldsymbol{n}}\times(\vec{\boldsymbol{a}}\times\vec{\boldsymbol{n}})]\cdot\hat{\boldsymbol{\sigma}}\}$$ ← $\vec{\boldsymbol{a}}\times(\vec{\boldsymbol{b}}\times\vec{\boldsymbol{c}}) = (\vec{\boldsymbol{a}}\cdot\vec{\boldsymbol{c}})\vec{\boldsymbol{b}} - (\vec{\boldsymbol{a}}\cdot\vec{\boldsymbol{b}})\vec{\boldsymbol{c}}$

$$= [\cos^2(\theta/2)\,\vec{\boldsymbol{a}} + \sin\theta\,(\vec{\boldsymbol{a}}\times\vec{\boldsymbol{n}}) + \sin^2(\theta/2)\,(\vec{\boldsymbol{a}}\cdot\vec{\boldsymbol{n}})\vec{\boldsymbol{n}}]\cdot\hat{\boldsymbol{\sigma}}$$

$$-\sin^2(\theta/2)\,[\underbrace{(\vec{\boldsymbol{n}}\cdot\vec{\boldsymbol{n}})}_{1}\vec{\boldsymbol{a}} - (\vec{\boldsymbol{a}}\cdot\vec{\boldsymbol{n}})\vec{\boldsymbol{n}}]\cdot\hat{\boldsymbol{\sigma}}$$

$$= [(\cos\theta)\vec{\boldsymbol{a}} + \sin\theta\,(\vec{\boldsymbol{a}}\times\vec{\boldsymbol{n}}) + (1-\cos\theta)(\vec{\boldsymbol{a}}\cdot\vec{\boldsymbol{n}})\vec{\boldsymbol{n}}]\cdot\hat{\boldsymbol{\sigma}}$$

$$= \{(\vec{\boldsymbol{a}}\cdot\vec{\boldsymbol{n}})\vec{\boldsymbol{n}} + \sin\theta\,(\vec{\boldsymbol{a}}\times\vec{\boldsymbol{n}}) + \cos\theta\,[\vec{\boldsymbol{a}} - (\vec{\boldsymbol{a}}\cdot\vec{\boldsymbol{n}})\vec{\boldsymbol{n}}]\}\cdot\hat{\boldsymbol{\sigma}} = \vec{\boldsymbol{a'}}\cdot\hat{\boldsymbol{\sigma}}. \qquad (35)$$

As shown in Fig. 5, $\vec{\boldsymbol{a'}}$ is the vector $\vec{\boldsymbol{a}}$ rotated around the unit-vector $\vec{\boldsymbol{n}}$ through the angle $\theta$ in accordance with the left-hand rule (with thumb along $\vec{\boldsymbol{n}}$). This is equivalent to a rotation of the coordinate system by $\theta$ around $\hat{\boldsymbol{n}}$ in the opposite direction (i.e., following the right-hand rule). In general, $\hat{u}$ acting on a state vector $|\psi\rangle$ corresponds to a right-handed rotation of the coordinate system around $\vec{\boldsymbol{n}}$ through the angle $\theta$. For the expected value $\langle\psi|\vec{\boldsymbol{a}}\cdot\hat{\boldsymbol{\sigma}}|\psi\rangle$ to remain the same before and after the rotation of coordinates, the operator $\vec{\boldsymbol{a}}\cdot\hat{\boldsymbol{\sigma}}$ must be transformed to $\hat{u}(\vec{\boldsymbol{a}}\cdot\hat{\boldsymbol{\sigma}})\hat{u}^\dagger$. Another way of seeing this is to note that an operator $\hat{O}$ with orthonormal eigenstates $|i\rangle$ and corresponding eigenvalues $\alpha_i$ can, in general, be written as $\hat{O} = \sum_i \alpha_i |i\rangle\langle i|$, hence the necessity of transforming it to $\hat{u}\hat{O}\hat{u}^\dagger$ upon a coordinate rotation.

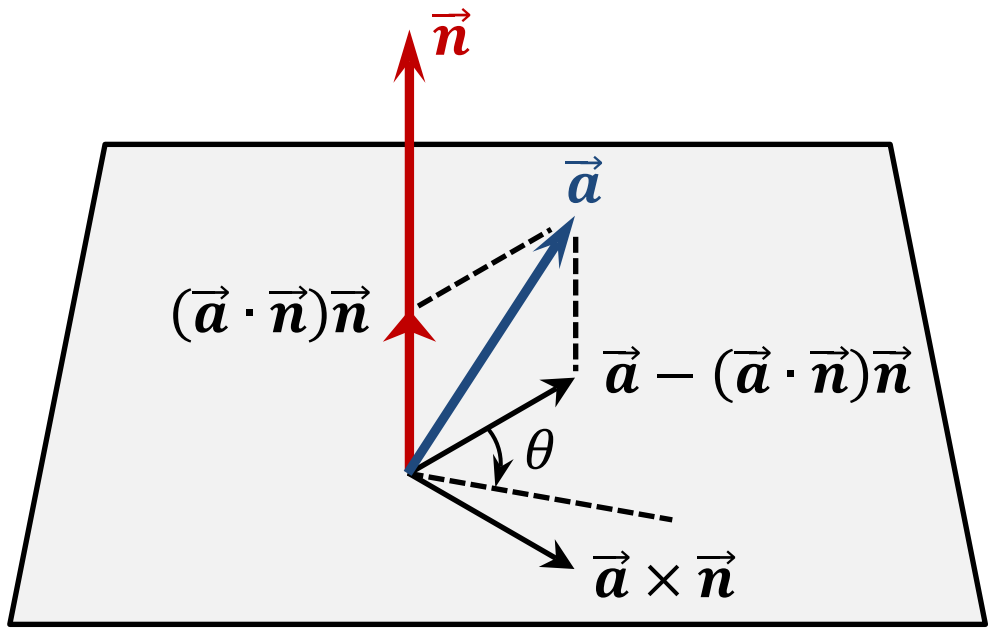

**Fig. 5**. The vector $(\vec{\boldsymbol{a}}\cdot\vec{\boldsymbol{n}})\vec{\boldsymbol{n}} + \sin\theta\,(\vec{\boldsymbol{a}}\times\vec{\boldsymbol{n}}) + \cos\theta\,[\vec{\boldsymbol{a}} - (\vec{\boldsymbol{a}}\cdot\vec{\boldsymbol{n}})\vec{\boldsymbol{n}}]$ arises from a left-hand-rule rotation of $\vec{\boldsymbol{a}}$ around the unit-vector $\vec{\boldsymbol{n}}$ through the angle $\theta$.

**8. Rotating the coordinates of a spin-½ particle via unitary transformations**. With reference to Fig.6, consider a unitary transformation $\hat{u}$ corresponding to a rotation around $z$ through the angle $\chi$, followed by tilting the rotated $x'y'z'$ system in such a way as to bring the $z'$-axis onto $z''$, whose polar and azimuthal coordinates within $x'y'z'$ are $\theta$ and $\phi$. (Note: Any $x''y''z''$ can be brought into alignment with $xyz$ by the same procedure in reverse.) We will have

$$\hat{u} = \exp[½\mathrm{i}\theta\hat{\sigma}\cdot(-\sin\phi\,\hat{\boldsymbol{x}} + \cos\phi\,\hat{\boldsymbol{y}})]\exp(½\mathrm{i}\chi\hat{\sigma}\cdot\hat{\boldsymbol{z}})$$

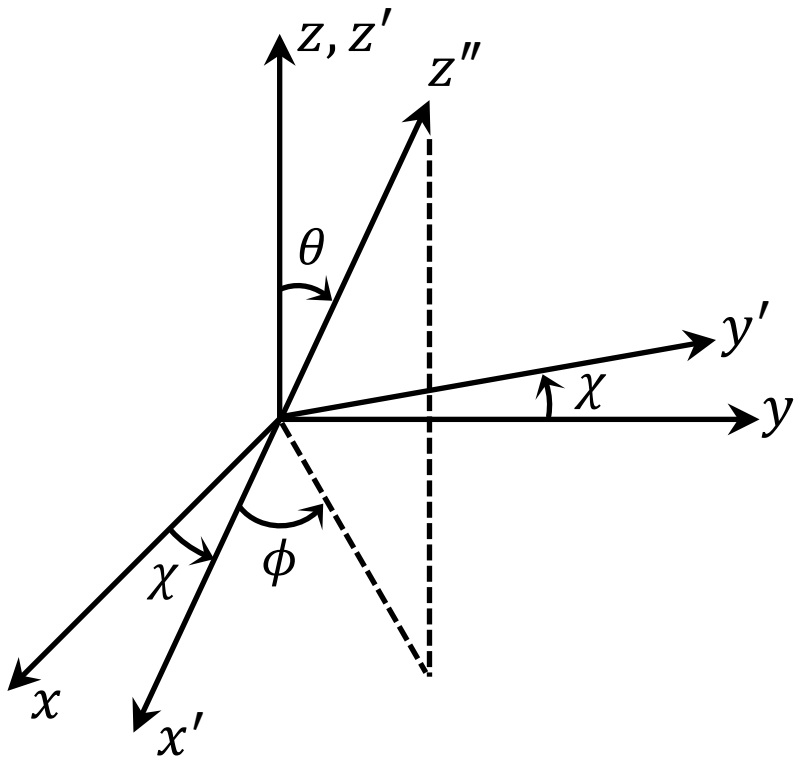

**Fig.6**. The Euler angles $(\theta, \phi, \chi)$ define a solid rotation of the $xyz$ coordinate system in three-dimensional space. A first rotation around the $z$-axis through the angle $\chi$ brings $xyz$ to $x'y'z'$. The $x'y'z'$ system is then tilted by $\theta$ to bring its $z'$-axis to $z''$ while keeping the projection of $z'$ onto the original $xy$-plane along a fixed direction (i.e., keeping the angle $\chi + \phi$ fixed as $z'$ moves away from $z$ toward $z''$). Note that $\chi$ is measured in the $xy$-plane from $x$ toward $y$, while $\phi$ is measured in the $x'y'$-plane from $x'$ toward $y'$. In this way, $x''y''z''$ can be made to have any desired orientation relative to $xyz$.

$\hat{\sigma} = \hat{\sigma}_x\hat{\boldsymbol{x}} + \hat{\sigma}_y\hat{\boldsymbol{y}} + \hat{\sigma}_z\hat{\boldsymbol{z}}$

$$= \left[\cos(\theta/2)\,\hat{\mathbb{I}} + \mathrm{i}\sin(\theta/2)\begin{pmatrix} 0 & -\mathrm{i}e^{-\mathrm{i}\phi} \\ \mathrm{i}e^{\mathrm{i}\phi} & 0 \end{pmatrix}\right]\left[\cos(\chi/2)\,\hat{\mathbb{I}} + \mathrm{i}\sin(\chi/2)\begin{pmatrix} 1 & 0 \\ 0 & -1 \end{pmatrix}\right]$$

$$= \cos(\theta/2)\cos(\chi/2)\,\hat{\mathbb{I}} + \mathrm{i}\cos(\theta/2)\sin(\chi/2)\begin{pmatrix} 1 & 0 \\ 0 & -1 \end{pmatrix}$$

$$+\mathrm{i}\sin(\theta/2)\cos(\chi/2)\begin{pmatrix} 0 & -\mathrm{i}e^{-\mathrm{i}\phi} \\ \mathrm{i}e^{\mathrm{i}\phi} & 0 \end{pmatrix} - \sin(\theta/2)\sin(\chi/2)\begin{pmatrix} 0 & \mathrm{i}e^{-\mathrm{i}\phi} \\ \mathrm{i}e^{\mathrm{i}\phi} & 0 \end{pmatrix}$$

$$= \begin{bmatrix} \cos(\theta/2)\,e^{\mathrm{i}\chi/2} & \sin(\theta/2)\,e^{-\mathrm{i}(\phi+½\chi)} \\ -\sin(\theta/2)\,e^{\mathrm{i}(\phi+½\chi)} & \cos(\theta/2)\,e^{-\mathrm{i}\chi/2} \end{bmatrix}. \quad (36)$$

Same as Feynman's Table 6-1, with $\theta = \alpha, \phi = -(90° + \gamma), \chi = \beta + \gamma$.[1]

For the spin state $|\psi\rangle$ of Eq.(28), rotating the coordinates through the Euler angles $(\theta, \phi, \chi)$ now yields

$$\begin{bmatrix} \cos(½\theta)\,e^{\mathrm{i}\chi/2} & \sin(½\theta)\,e^{-\mathrm{i}(\phi+½\chi)} \\ -\sin(½\theta)\,e^{\mathrm{i}(\phi+½\chi)} & \cos(½\theta)\,e^{-\mathrm{i}\chi/2} \end{bmatrix}\begin{bmatrix} \cos(½\theta_0) \\ \sin(½\theta_0)\,e^{\mathrm{i}\phi_0} \end{bmatrix}$$

$$= e^{\mathrm{i}\chi/2}\begin{bmatrix} \cos(½\theta)\cos(½\theta_0) + \sin(½\theta)\sin(½\theta_0)\,e^{\mathrm{i}(\phi_0-\phi-\chi)} \\ -\sin(½\theta)\cos(½\theta_0)\,e^{\mathrm{i}\phi} + \cos(½\theta)\sin(½\theta_0)\,e^{\mathrm{i}(\phi_0-\chi)} \end{bmatrix}. \quad (37)$$

Thus, when $(\theta_0, \phi_0) = (0°, 0°)$, the state ends up as $\cos(½\theta)\,|\uparrow\rangle + \sin(½\theta)\,e^{\mathrm{i}(\phi+\pi)}|\downarrow\rangle$. Similarly, when $(\theta_0, \phi_0) = (180°, 0°)$, the state after the rotation becomes $\cos[½(\pi-\theta)]\,|\uparrow\rangle + \sin[½(\pi-\theta)]\,e^{\mathrm{i}\phi}|\downarrow\rangle$. Consider now the unitary rotation operator around $\hat{\boldsymbol{y}}$ through $\theta$, namely,

$$\hat{u} = \cos(½\theta)\,\hat{\mathbb{I}} + \mathrm{i}\sin(½\theta)\,(\hat{\sigma}\cdot\hat{\boldsymbol{y}}) = \begin{bmatrix} \cos(½\theta) & \sin(½\theta) \\ -\sin(½\theta) & \cos(½\theta) \end{bmatrix}. \quad (38)$$

Same as Eq.(36), with $\phi = \chi = 0$

Let the electrons in the state $|\psi\rangle = (|\uparrow\rangle_1|\downarrow\rangle_2 - |\downarrow\rangle_1|\uparrow\rangle_2)/\sqrt{2}$ of a maximally entangled pair of electrons pass through Stern-Gerlach analyzers that have been rotated around $\hat{\boldsymbol{y}}$ through the angles $\theta_1$ and $\theta_2$, respectively. In the new (rotated) basis, the state $|\psi\rangle$ is expressed as

$$|\psi\rangle = \tfrac{1}{\sqrt{2}}\{[\cos(½\theta_1)\,|\uparrow\rangle - \sin(½\theta_1)\,|\downarrow\rangle]_1[\sin(½\theta_2)\,|\uparrow\rangle + \cos(½\theta_2)\,|\downarrow\rangle]_2$$
$$-[\sin(½\theta_1)\,|\uparrow\rangle + \cos(½\theta_1)\,|\downarrow\rangle]_1[\cos(½\theta_2)\,|\uparrow\rangle - \sin(½\theta_2)\,|\downarrow\rangle]_2\}$$
$$= \tfrac{1}{\sqrt{2}}\{\sin[½(\theta_2-\theta_1)]\,|\uparrow\rangle_1|\uparrow\rangle_2 + \cos[½(\theta_2-\theta_1)]|\uparrow\rangle_1|\downarrow\rangle_2$$
$$-\cos[½(\theta_2-\theta_1)]\,|\downarrow\rangle_1|\uparrow\rangle_2 + \sin[½(\theta_2-\theta_1)]\,|\downarrow\rangle_1|\downarrow\rangle_2\}. \quad (39)$$

Clearly, when $\theta_1 = \theta_2$, the electron pair remains in its initial state, and when $\theta_1 \neq \theta_2$, the two spins remain antiparallel with a probability of $\cos^2[½(\theta_2-\theta_1)]$. If each pair of electrons carried instructions as to how to emerge from Stern-Gerlach analyzers set at different angles $\theta$, the instruction set for each member of the pair would have to be the exact opposite of that for the other member for any specific $\theta$. Assuming each analyzer is randomly set to $\theta = 0°$, $120°$, or $240°$, the instruction sets must belong to one of the eight possible categories listed in Table 2. In the case of instruction sets 1 and 8, all measurements at the output of the two stern-Gerlach analyzers would yield antiparallel spins. For all the remaining instruction sets, the electrons will emerge with antiparallel spins in one-third of the experiments in which $\theta_1 \neq \theta_2$. It is thus clear that the overall chances of observing antiparallel spins at the exit ports of the analyzers will be greater than 33.3%. In contrast, since $\cos^2[½(\theta_2 - \theta_1)] = \cos^2(60°) = 0.25$, the quantum mechanical expression in Eq.(39) predicts that only 25% of the experiments would result in antiparallel spins. Once again, quantum mechanics is seen to be incompatible with LHVTs.

| $\theta$ ↓ \ instruction set → | 1 | 2 | 3 | 4 | 5 | 6 | 7 | 8 |
|---|---|---|---|---|---|---|---|---|
| 0° | ↑↓ | ↓↑ | ↑↓ | ↓↑ | ↑↓ | ↓↑ | ↑↓ | ↓↑ |
| 120° | ↑↓ | ↑↓ | ↓↑ | ↓↑ | ↑↓ | ↑↓ | ↓↑ | ↓↑ |
| 240° | ↑↓ | ↑↓ | ↑↓ | ↑↓ | ↓↑ | ↓↑ | ↓↑ | ↓↑ |

**Table 2**. In experiments where each Stern-Gerlach analyzer is rotated around the $y$-axis and set randomly and independently of the other analyzer to $\theta = 0°$, 120°, or 240°, each pair of entangled electrons must carry one of eight possible instruction sets corresponding to the three analyzer settings that they might encounter. The arrow on the left-hand side in each depicted pair tells the first electron how to emerge, while the arrow on the right-hand side gives the second electron its instructions. In cases 1 and 8, the two electrons will always be observed with antiparallel spins. In all the other cases, antiparallel spins will be observed, on average, in one third of the experiments in which $\theta_1 \neq \theta_2$.

**Digression**. A photon propagating along the $z$-axis is, in general, in the superposition state $|\psi\rangle = \alpha_1|\uparrow\rangle + \alpha_2|\downarrow\rangle = [(\alpha_1 + \alpha_2)|x\rangle + \mathrm{i}(\alpha_1 - \alpha_2)|y\rangle]/\sqrt{2}$. Rotating the $xyz$ coordinate system through the angle $\chi$ around $\hat{\boldsymbol{z}}$ requires a different unitary transformation than that for spin-½ particles (compare Feynman's Tables 17-1 and 17-3[1]), namely,

$$\hat{u}_{\text{ph}} = \cos\chi\,\hat{\mathbb{I}} + \mathrm{i}\sin\chi\,(\hat{\sigma}\cdot\hat{\boldsymbol{z}}) = \begin{pmatrix} e^{\mathrm{i}\chi} & 0 \\ 0 & e^{-\mathrm{i}\chi} \end{pmatrix}. \tag{40}$$

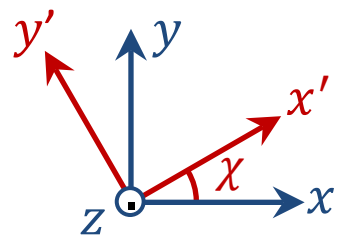

The state of the photon in the rotated coordinates is thus written as follows:

$$\hat{u}_{\text{ph}}|\psi\rangle = \alpha_1 e^{\mathrm{i}\chi}|\uparrow\rangle + \alpha_2 e^{-\mathrm{i}\chi}|\downarrow\rangle = \tfrac{1}{\sqrt{2}}[\alpha_1 e^{\mathrm{i}\chi}(|x'\rangle + \mathrm{i}|y'\rangle) + \alpha_2 e^{-\mathrm{i}\chi}(|x'\rangle - \mathrm{i}|y'\rangle)]$$

$$= \tfrac{1}{\sqrt{2}}[(\alpha_1 + \alpha_2)(\cos\chi\,|x'\rangle - \sin\chi\,|y'\rangle) + \mathrm{i}(\alpha_1 - \alpha_2)(\sin\chi\,|x'\rangle + \cos\chi\,|y'\rangle)]. \tag{41}$$

This is what one would obtain by expressing $|x\rangle$ and $|y\rangle$ in the original expression of $|\psi\rangle$ in terms of $|x'\rangle$ and $|y'\rangle$. Note, however, that $\hat{u}_{\text{ph}}$ will *not* act correctly on the photon's state if $|\psi\rangle$ is expressed in linearly-polarized $|x\rangle, |y\rangle$ basis.

**Digression**. The rotation operator of Eq.(36) can be used to arrive at the spin angular momentum operators $\hat{S}_x, \hat{S}_y, \hat{S}_z$.[1] Noting that rotation around the $x$-axis through the small angle $\varepsilon$ corresponds to $\theta = \varepsilon$, $\phi = -90°$, $\chi = 0°$, we write

$$\hat{R}_x(\varepsilon) = \begin{bmatrix} \cos(\varepsilon/2) & \mathrm{i}\sin(\varepsilon/2) \\ \mathrm{i}\sin(\varepsilon/2) & \cos(\varepsilon/2) \end{bmatrix} \cong \begin{pmatrix} 1 & ½\mathrm{i}\varepsilon \\ ½\mathrm{i}\varepsilon & 1 \end{pmatrix} = \hat{\mathbb{I}} + \mathrm{i}(\varepsilon/\hbar)\hat{S}_x \quad \rightarrow \quad \hat{S}_x = ½\hbar\begin{pmatrix} 0 & 1 \\ 1 & 0 \end{pmatrix}. \tag{42a}$$

Similarly, given that a rotation around $y$ through the small angle $\varepsilon$ corresponds to $\theta = \varepsilon$, $\phi = \chi = 0°$, we will have

$$\hat{R}_y(\varepsilon) = \begin{bmatrix} \cos(\varepsilon/2) & \sin(\varepsilon/2) \\ -\sin(\varepsilon/2) & \cos(\varepsilon/2) \end{bmatrix} \cong \begin{pmatrix} 1 & ½\varepsilon \\ -½\varepsilon & 1 \end{pmatrix} = \hat{\mathbb{I}} + \mathrm{i}(\varepsilon/\hbar)\hat{S}_y \quad \rightarrow \quad \hat{S}_y = ½\hbar\begin{pmatrix} 0 & -\mathrm{i} \\ \mathrm{i} & 0 \end{pmatrix}. \tag{42b}$$

Finally, since a rotation around $z$ through the small angle $\varepsilon$ corresponds to $\theta = \phi = 0°$, $\chi = \varepsilon$, we will have

$$\hat{R}_z(\varepsilon) = \begin{bmatrix} e^{\mathrm{i}\varepsilon/2} & 0 \\ 0 & e^{-\mathrm{i}\varepsilon/2} \end{bmatrix} \cong \begin{pmatrix} 1 + ½\mathrm{i}\varepsilon & 0 \\ 0 & 1 - ½\mathrm{i}\varepsilon \end{pmatrix} = \hat{\mathbb{I}} + \mathrm{i}(\varepsilon/\hbar)\hat{S}_z \quad \rightarrow \quad \hat{S}_z = ½\hbar\begin{pmatrix} 1 & 0 \\ 0 & -1 \end{pmatrix}. \tag{42c}$$

**9. A Bell inequality**. In our discussion of hidden variables in Sec.2, we assumed that each entangled pair of photons came equipped with an instruction set that told photon 1 (2) to pass or not pass through analyzer 1 (2) if the analyzer's angle was set to $\theta_1$ ($\theta_2$). We also assumed the instructions varied randomly from one photon pair to the next, so that, upon emerging from the source, each pair received a randomly drawn instruction set from a pile of conceivable such sets.

Let us now formalize the procedure by introducing the hidden random variable $\zeta$ which has some (unknown) probability distribution $p(\zeta)$ that satisfies the conventional requirements for such distributions; that is, $p(\zeta) \geq 0$ and $\int_{-\infty}^{\infty} p(\zeta)\mathrm{d}\zeta = 1$. Each instruction set contains two graphs, $\Omega_1(\theta_1)$ for use by photon 1, and $\Omega_2(\theta_2)$ for photon 2. In each graph, the range of $\theta$ is $[0, 90°)$ and $\Omega(\theta)$ is $+1$ for some values of $\theta$ and $-1$ for all the remaining values, so that the corresponding photon, upon encountering an analyzer oriented at $\theta$, would pass through to the detector if $\Omega(\theta) = +1$, but refuse to go through if $\Omega(\theta) = -1$. To acknowledge the random nature of these graphs, we denote their dependence on the hidden variable $\zeta$ by writing them as $\Omega_1(\theta_1; \zeta)$ and $\Omega_2(\theta_2; \zeta)$. Also, considering that each photon is known to have a 50% chance of being transmitted (and 50% chance of not being transmitted) to a detector irrespective of the corresponding analyzer's orientation angle, one may assume that $\int_{-\infty}^{\infty} p(\zeta)\Omega_1(\theta_1; \zeta)\mathrm{d}\zeta = \int_{-\infty}^{\infty} p(\zeta)\Omega_2(\theta_2; \zeta)\mathrm{d}\zeta = 0$, although this assumption has no effect on the arguments that follow.

We shall be interested in the correlations between the outputs of detectors 1 and 2 depicted in Fig.1. Specifically, we aim to keep track of the product $\Omega_1(\theta_1; \zeta)\Omega_2(\theta_2; \zeta)$ after each run of the experiment, recognizing that the product will be $+1$ if both photons reach their respective detectors or both fail to do so; the product will be $-1$ if only one of the photons arrives at its designated detector. Quantum mechanics tells us that the photon pair behaves in accordance with Eq.(3); that is: (i) both photons reach their respective detectors with probability $\tfrac{1}{2}\cos^2(\theta_1 - \theta_2)$, (ii) neither photon reaches its detector with probability $\tfrac{1}{2}\cos^2(\theta_1 - \theta_2)$, (iii) only detector 1 picks up a photon with probability $\tfrac{1}{2}\sin^2(\theta_1 - \theta_2)$, and (iv) only detector 2 picks up a photon with probability $\tfrac{1}{2}\sin^2(\theta_1 - \theta_2)$. All in all, presuming the validity of quantum mechanical predictions, the expected value (identified with an overbar) of the product $\Omega_1(\theta_1; \zeta)\Omega_2(\theta_2; \zeta)$ must be

$$\overline{\Omega_1(\theta_1; \zeta)\Omega_2(\theta_2; \zeta)} = \cos^2(\theta_1 - \theta_2) - \sin^2(\theta_1 - \theta_2) = \cos[2(\theta_1 - \theta_2)]. \tag{43}$$

A similar argument can be advanced for the entangled pair of electrons examined in Sec.8, with each electron emerging from its analyzer in either spin-up $|\uparrow\rangle$ or spin-down $|\downarrow\rangle$ state depending on whether the corresponding $\Omega(\theta; \zeta)$ is $+1$ or $-1$. For this case, Eq.(39) yields the expected value of the correlation between the joint detection outcomes as $\overline{\Omega_1(\theta_1; \zeta)\Omega_2(\theta_2; \zeta)} = -\cos(\theta_1 - \theta_2)$.

Next, let us consider a large number of experiments in which one of two values, $\theta_1$ or $\theta_1'$, is randomly chosen for the first analyzer, and also one of two values, $\theta_2$ or $\theta_2'$, is randomly chosen for the second analyzer. We thus have four different settings of the two analyzers for each conducted experiment, namely, $(\theta_1, \theta_2)$, $(\theta_1, \theta_2')$, $(\theta_1', \theta_2)$, and $(\theta_1', \theta_2')$ — each chosen randomly and with equal probability. We keep track of the correlations $\Omega_1\Omega_2$ of the outcomes of these experiments, each tagged with the corresponding pair of the orientation angles of the analyzers.

Returning now to the instruction sets presumed to be carried by our entangled particle pairs, let us consider the following combination of the possible experimental outcomes according to LHVTs:

$$\Gamma(\theta_1, \theta_1', \theta_2, \theta_2'; \zeta) = \Omega_1(\theta_1; \zeta)\Omega_2(\theta_2; \zeta) + \Omega_1(\theta_1; \zeta)\Omega_2(\theta_2'; \zeta) + \Omega_1(\theta_1'; \zeta)\Omega_2(\theta_2; \zeta) - \Omega_1(\theta_1'; \zeta)\Omega_2(\theta_2'; \zeta). \tag{44}$$

The four terms on the right-hand side of Eq.(44) can be rearranged in the following way:

$$\Gamma(\theta_1, \theta_1', \theta_2, \theta_2'; \zeta) = \Omega_1(\theta_1; \zeta)[\Omega_2(\theta_2; \zeta) + \Omega_2(\theta_2'; \zeta)] + \Omega_1(\theta_1'; \zeta)[\Omega_2(\theta_2; \zeta) - \Omega_2(\theta_2'; \zeta)]. \tag{45}$$

Given that $\Omega_1$ and $\Omega_2$ assume only $\pm 1$ values, it is not difficult to see from Eq.(45) that, for any randomly chosen $\zeta$, the value of $\Gamma$ will be either $+2$ or $-2$. Thus, the expected value of $\Gamma$ must satisfy the inequality

$$-2 \le \overline{\Gamma(\theta_1,\theta_1',\theta_2,\theta_2';\zeta)} = \int_{-\infty}^{\infty} p(\zeta)\Gamma(\theta_1,\theta_1',\theta_2,\theta_2';\zeta)\mathrm{d}\zeta \le 2. \quad (46)$$

Equation (44) in conjunction with inequality (46) leads to the inescapable conclusion that, according to LHVTs, the expected values of the aforementioned four correlation measurements must satisfy the following Bell inequality:[§]

$$-2 \le \overline{\Omega_1(\theta_1;\zeta)\Omega_2(\theta_2;\zeta)} + \overline{\Omega_1(\theta_1;\zeta)\Omega_2(\theta_2';\zeta)} + \overline{\Omega_1(\theta_1';\zeta)\Omega_2(\theta_2;\zeta)} - \overline{\Omega_1(\theta_1';\zeta)\Omega_2(\theta_2';\zeta)} \le 2. \quad (47)$$

As it turns out, quantum mechanical predictions violate Bell's inequality in spectacular ways. As a simple example, consider the case of an entangled pair of photons with analyzer settings at $\theta_1 = 45°$, $\theta_1' = 90°$, $\theta_2 = 67.5°$, $\theta_2' = 22.5°$. From Eq.(43), one expects to find $\overline{\Omega_1(\theta_1;\zeta)\Omega_2(\theta_2;\zeta)} = \overline{\Omega_1(\theta_1;\zeta)\Omega_2(\theta_2';\zeta)} = \overline{\Omega_1(\theta_1';\zeta)\Omega_2(\theta_2;\zeta)} = \sqrt{2}/2$ and $\overline{\Omega_1(\theta_1';\zeta)\Omega_2(\theta_2';\zeta)} = -\sqrt{2}/2$. The combination of these four expected values at $2\sqrt{2}$ is far outside the $[-2,2]$ range predicted by LHVTs. Once again, quantum mechanics is found to be incompatible with the existence of local hidden variables.

**10. Singlet and triplet states of a pair of spin-½ particles**. The state $|\psi_1\rangle = (|\uparrow\downarrow\rangle - |\downarrow\uparrow\rangle)/\sqrt{2}$ is known as the singlet, while the states $|\psi_2\rangle = |\uparrow\uparrow\rangle$, $|\psi_3\rangle = (|\uparrow\downarrow\rangle + |\downarrow\uparrow\rangle)/\sqrt{2}$, and $|\psi_4\rangle = |\downarrow\downarrow\rangle$ are known as the triplet states.[1] These four states, being individually normalized and mutually orthogonal, form a basis for our two-spin system. They are also the eigenstates of the operators $\hat{S}_z = ½\hbar(\hat{\sigma}_{z1} + \hat{\sigma}_{z2})$ and $\hat{S}^2 = \hat{S}_x^2 + \hat{S}_y^2 + \hat{S}_z^2 = ¼\hbar^2[(\hat{\sigma}_{x1} + \hat{\sigma}_{x2})^2 + (\hat{\sigma}_{y1} + \hat{\sigma}_{y2})^2 + (\hat{\sigma}_{z1} + \hat{\sigma}_{z2})^2] = ½\hbar^2(3\hat{\mathbb{I}} + \hat{\sigma}_{x1}\hat{\sigma}_{x2} + \hat{\sigma}_{y1}\hat{\sigma}_{y2} + \hat{\sigma}_{z1}\hat{\sigma}_{z2})$.[**] The triplet states $|\psi_2\rangle$, $|\psi_3\rangle$, $|\psi_4\rangle$ have the common eigenvalue of $2\hbar^2$ for $\hat{S}^2$, and respective eigenvalues of $+\hbar$, 0, $-\hbar$ for $\hat{S}_z$. The eigenvalue of $\hat{S}^2$ is commonly written as $S(S+1)\hbar^2$. Thus, each and every one of the triplet states has $S = 1$. In contrast, the eigenvalues of $\hat{S}^2$ and $\hat{S}_z$ for the singlet state $|\psi_1\rangle$ are $S = S_z = 0$, corresponding to a state of zero net angular momentum.

Using the triplet states as a model for a spin-1 system, we now derive the corresponding rotation operator by invoking the rotation matrix of Eq.(36) for the underlying pair of spin-½ particles. Rotating the $xyz$ coordinate system through the Euler angles $(\theta,\phi,\chi)$ results in the following transformations for each of the spin-½ particles:

$$|\uparrow\rangle \;\rightarrow\; \cos(\theta/2)\, e^{\mathrm{i}\chi/2}\, |\uparrow\rangle - \sin(\theta/2)\, e^{\mathrm{i}(\phi+½\chi)}|\downarrow\rangle, \quad (48a)$$

$$|\downarrow\rangle \;\rightarrow\; \sin(\theta/2)\, e^{-\mathrm{i}(\phi+½\chi)}\, |\uparrow\rangle_2 + \cos(\theta/2)\, e^{-\mathrm{i}\chi/2}|\downarrow\rangle_2. \quad (48b)$$

Applying the above transformations to the particle pair that comprises the spin-1 states yields

$$\begin{aligned} |\uparrow\rangle_1|\uparrow\rangle_2 &= \left[\cos(\theta/2)\, e^{\mathrm{i}\chi/2}\, |\uparrow\rangle_1 - \sin(\theta/2)\, e^{\mathrm{i}(\phi+½\chi)}|\downarrow\rangle_1\right] \\ &\quad \times \left[\cos(\theta/2)\, e^{\mathrm{i}\chi/2}\, |\uparrow\rangle_2 - \sin(\theta/2)\, e^{\mathrm{i}(\phi+½\chi)}|\downarrow\rangle_2\right] \\ &= ½(1+\cos\theta)e^{\mathrm{i}\chi}|\uparrow\rangle_1|\uparrow\rangle_2 - ½\sin\theta\, e^{\mathrm{i}(\phi+\chi)}(|\uparrow\rangle_1|\downarrow\rangle_2 + |\downarrow\rangle_1|\uparrow\rangle_2) \\ &\quad +½(1-\cos\theta)e^{\mathrm{i}(2\phi+\chi)}|\downarrow\rangle_1|\downarrow\rangle_2. \end{aligned} \quad (49a)$$

[§] Inequality (47) is in fact a generalized (and more powerful) version of the original Bell inequality, often referred to as the Clauser-Holt-Horne-Shimony (CHHS) Bell inequality.[13] A noteworthy feature of this particular version is that it does *not* require $\overline{\Omega_1(\theta_1;\zeta)\Omega_2(\theta_2;\zeta)}$ to equal 1 when $\theta_1 = \theta_2$.

[**] $\hat{\boldsymbol{\sigma}} \cdot \hat{\boldsymbol{\sigma}} = \hat{\sigma}_x^2 + \hat{\sigma}_y^2 + \hat{\sigma}_z^2 = \begin{pmatrix} 0 & 1 \\ 1 & 0 \end{pmatrix}\begin{pmatrix} 0 & 1 \\ 1 & 0 \end{pmatrix} + \begin{pmatrix} 0 & -\mathrm{i} \\ \mathrm{i} & 0 \end{pmatrix}\begin{pmatrix} 0 & -\mathrm{i} \\ \mathrm{i} & 0 \end{pmatrix} + \begin{pmatrix} 1 & 0 \\ 0 & -1 \end{pmatrix}\begin{pmatrix} 1 & 0 \\ 0 & -1 \end{pmatrix} = 3\hat{\mathbb{I}}$.

$$\tfrac{1}{\sqrt{2}}(|\uparrow\rangle_1|\downarrow\rangle_2+|\downarrow\rangle_1|\uparrow\rangle_2)=\tfrac{1}{\sqrt{2}}\left[\cos(\theta/2)\,e^{\mathrm{i}\chi/2}\,|\uparrow\rangle_1-\sin(\theta/2)\,e^{\mathrm{i}(\phi+½\chi)}|\downarrow\rangle_1\right]$$

$$\times\left[\sin(\theta/2)\,e^{-\mathrm{i}(\phi+½\chi)}\,|\uparrow\rangle_2+\cos(\theta/2)\,e^{-\mathrm{i}\chi/2}|\downarrow\rangle_2\right]$$

$$+\tfrac{1}{\sqrt{2}}\left[\sin(\theta/2)\,e^{-\mathrm{i}(\phi+½\chi)}\,|\uparrow\rangle_1+\cos(\theta/2)\,e^{-\mathrm{i}\chi/2}|\downarrow\rangle_1\right]$$

$$\times\left[\cos(\theta/2)\,e^{\mathrm{i}\chi/2}\,|\uparrow\rangle_2-\sin(\theta/2)\,e^{\mathrm{i}(\phi+½\chi)}|\downarrow\rangle_2\right]$$

$$=\tfrac{1}{\sqrt{2}}\sin\theta\,e^{-\mathrm{i}\phi}|\uparrow\rangle_1|\uparrow\rangle_2+\tfrac{1}{\sqrt{2}}\cos\theta\,(|\uparrow\rangle_1|\downarrow\rangle_2+|\downarrow\rangle_1|\uparrow\rangle_2)-\tfrac{1}{\sqrt{2}}\sin\theta\,e^{\mathrm{i}\phi}|\downarrow\rangle_1|\downarrow\rangle_2. \quad (49b)$$

$$|\downarrow\rangle_1|\downarrow\rangle_2=\left[\sin(\theta/2)\,e^{-\mathrm{i}(\phi+½\chi)}\,|\uparrow\rangle_1+\cos(\theta/2)\,e^{-\mathrm{i}\chi/2}|\downarrow\rangle_1\right]$$

$$\times\left[\sin(\theta/2)\,e^{-\mathrm{i}(\phi+½\chi)}\,|\uparrow\rangle_2+\cos(\theta/2)\,e^{-\mathrm{i}\chi/2}|\downarrow\rangle_2\right]$$

$$=½(1-\cos\theta)e^{-\mathrm{i}(2\phi+\chi)}|\uparrow\rangle_1|\uparrow\rangle_2+½\sin\theta\,e^{-\mathrm{i}(\phi+\chi)}(|\uparrow\rangle_1|\downarrow\rangle_2+|\downarrow\rangle_1|\uparrow\rangle_2)$$

$$+½(1+\cos\theta)e^{-\mathrm{i}\chi}|\downarrow\rangle_1|\downarrow\rangle_2. \quad (49c)$$

Combining these results and ordering them in terms of the projection of the spin-1 particle along the $z$-axis, i.e., the $S_z=\hbar$ state $|\uparrow\uparrow\rangle$, the $S_z=0$ state $(|\uparrow\uparrow\rangle+|\downarrow\downarrow\rangle)/\sqrt{2}$, and the $S_z=-\hbar$ state $|\downarrow\downarrow\rangle$, we arrive at the following $3\times3$ rotation matrix:

$$\hat{u}=\begin{bmatrix} ½(1+\cos\theta)e^{\mathrm{i}\chi} & \tfrac{1}{\sqrt{2}}\sin\theta\,e^{-\mathrm{i}\phi} & ½(1-\cos\theta)e^{-\mathrm{i}(2\phi+\chi)} \\ -\tfrac{1}{\sqrt{2}}\sin\theta\,e^{\mathrm{i}(\phi+\chi)} & \cos\theta & \tfrac{1}{\sqrt{2}}\sin\theta\,e^{-\mathrm{i}(\phi+\chi)} \\ ½(1-\cos\theta)e^{\mathrm{i}(2\phi+\chi)} & -\tfrac{1}{\sqrt{2}}\sin\theta\,e^{\mathrm{i}\phi} & ½(1+\cos\theta)e^{-\mathrm{i}\chi} \end{bmatrix}. \quad (50)$$

**11. The angular momentum operators $\hat{J}_x, \hat{J}_y, \hat{J}_z$ for a spin-1 particle**. Rotating the coordinate system around $x$ through the small angle $\varepsilon$ (corresponding to $\theta=\varepsilon$, $\phi=-90°$, $\chi=0°$) yields

$$\hat{R}_x(\varepsilon)=\begin{pmatrix} 1-¼\varepsilon^2 & \mathrm{i}\varepsilon/\sqrt{2} & -¼\varepsilon^2 \\ \mathrm{i}\varepsilon/\sqrt{2} & 1-½\varepsilon^2 & \mathrm{i}\varepsilon/\sqrt{2} \\ -¼\varepsilon^2 & \mathrm{i}\varepsilon/\sqrt{2} & 1-¼\varepsilon^2 \end{pmatrix}=1+\mathrm{i}(\varepsilon/\hbar)\hat{J}_x \;\rightarrow\; \hat{J}_x=\tfrac{\hbar}{\sqrt{2}}\begin{pmatrix} 0 & 1 & 0 \\ 1 & 0 & 1 \\ 0 & 1 & 0 \end{pmatrix}. \quad (51)$$

Next, we rotate the coordinates around the $y$-axis through the small angle $\varepsilon$ (corresponding to $\theta=\varepsilon$, $\phi=\chi=0°$) to obtain

$$\hat{R}_y(\varepsilon)=\begin{pmatrix} 1-¼\varepsilon^2 & \varepsilon/\sqrt{2} & ¼\varepsilon^2 \\ -\varepsilon/\sqrt{2} & 1-½\varepsilon^2 & \varepsilon/\sqrt{2} \\ ¼\varepsilon^2 & -\varepsilon/\sqrt{2} & 1-¼\varepsilon^2 \end{pmatrix}=1+\mathrm{i}(\varepsilon/\hbar)\hat{J}_y \;\rightarrow\; \hat{J}_y=\tfrac{\hbar}{\sqrt{2}}\begin{pmatrix} 0 & -\mathrm{i} & 0 \\ \mathrm{i} & 0 & -\mathrm{i} \\ 0 & \mathrm{i} & 0 \end{pmatrix}. \quad (52)$$

Finally, rotating the coordinates around the $z$-axis through the small angle $\varepsilon$ (corresponding to $\theta=\phi=0°$, $\chi=\varepsilon$) yields

$$\hat{R}_z(\varepsilon)=\begin{pmatrix} e^{\mathrm{i}\varepsilon} & 0 & 0 \\ 0 & 1 & 0 \\ 0 & 0 & e^{-\mathrm{i}\varepsilon} \end{pmatrix}=1+\mathrm{i}(\varepsilon/\hbar)\hat{J}_z \qquad\rightarrow\qquad \hat{J}_z=\hbar\begin{pmatrix} 1 & 0 & 0 \\ 0 & 0 & 0 \\ 0 & 0 & -1 \end{pmatrix}. \quad (53)$$

Squaring the above operators, we find

$$\hat{J}_x^2 = ½\hbar^2 \begin{pmatrix} 0 & 1 & 0 \\ 1 & 0 & 1 \\ 0 & 1 & 0 \end{pmatrix} \begin{pmatrix} 0 & 1 & 0 \\ 1 & 0 & 1 \\ 0 & 1 & 0 \end{pmatrix} = \hbar^2 \begin{pmatrix} ½ & 0 & ½ \\ 0 & 1 & 0 \\ ½ & 0 & ½ \end{pmatrix}. \tag{54}$$

$$\hat{J}_y^2 = ½\hbar^2 \begin{pmatrix} 0 & -\mathrm{i} & 0 \\ \mathrm{i} & 0 & -\mathrm{i} \\ 0 & \mathrm{i} & 0 \end{pmatrix} \begin{pmatrix} 0 & -\mathrm{i} & 0 \\ \mathrm{i} & 0 & -\mathrm{i} \\ 0 & \mathrm{i} & 0 \end{pmatrix} = \hbar^2 \begin{pmatrix} ½ & 0 & -½ \\ 0 & 1 & 0 \\ -½ & 0 & ½ \end{pmatrix}. \tag{55}$$

$$\hat{J}_z^2 = \hbar^2 \begin{pmatrix} 1 & 0 & 0 \\ 0 & 0 & 0 \\ 0 & 0 & -1 \end{pmatrix} \begin{pmatrix} 1 & 0 & 0 \\ 0 & 0 & 0 \\ 0 & 0 & -1 \end{pmatrix} = \hbar^2 \begin{pmatrix} 1 & 0 & 0 \\ 0 & 0 & 0 \\ 0 & 0 & 1 \end{pmatrix}. \tag{56}$$

Consequently, $\hat{\boldsymbol{J}}^2 = \hat{J}_x^2 + \hat{J}_y^2 + \hat{J}_z^2 = 2\hbar^2\hat{\mathbb{I}}$. Consider now a two-particle system consisting of a spin-1 particle $a$ and a spin-½ particle $b$. The state $|\Uparrow\rangle_a|\uparrow\rangle_b$ of the system is an eigenstate of $\hat{J}_{z,a} + \hat{S}_{z,b}$ with the eigenvalue $\hbar + ½\hbar = 3\hbar/2$. As for the total angular momentum, we have

$$\begin{aligned}(\hat{\boldsymbol{J}}_a + \hat{\boldsymbol{S}}_b)\cdot(\hat{\boldsymbol{J}}_a + \hat{\boldsymbol{S}}_b)|\Uparrow\rangle_a|\uparrow\rangle_b &= [(\hat{J}_x + \hat{S}_x)^2 + (\hat{J}_y + \hat{S}_y)^2 + (\hat{J}_z + \hat{S}_z)^2]\,|\Uparrow\rangle_a|\uparrow\rangle_b \\ &= [\hat{\boldsymbol{J}}_a^2 + \hat{\boldsymbol{S}}_b^2 + 2(\hat{J}_{x,a}\hat{S}_{x,b} + \hat{J}_{y,a}\hat{S}_{y,b} + \hat{J}_{z,a}\hat{S}_{z,b})]\,|\Uparrow\rangle_a|\uparrow\rangle_b \\ &= (2\hbar^2\hat{\mathbb{I}} + ¾\hbar^2\hat{\mathbb{I}})|\Uparrow\rangle_a|\uparrow\rangle_b + 2(\hbar/\sqrt{2})(½\hbar)|0\rangle_a|\downarrow\rangle_b \\ &\quad + 2(\mathrm{i}\hbar/\sqrt{2})(½\mathrm{i}\hbar)|0\rangle_a|\downarrow\rangle_b + 2\hbar(½\hbar)\,|\Uparrow\rangle_a|\uparrow\rangle_b \\ &= \tfrac{3}{2}\left(1 + \tfrac{3}{2}\right)\hbar^2|\Uparrow\rangle_a|\uparrow\rangle_b. \end{aligned} \tag{57}$$

The states of particle $a$ relative to the $z$-axis are $|\Uparrow\rangle$, $|0\rangle$, and $|\Downarrow\rangle$.

A similar treatment shows $|\Downarrow\rangle_a|\downarrow\rangle_b$ to be an eigenstate of $\hat{J}_{z,a} + \hat{S}_{z,b}$ with the eigenvalue $-3\hbar/2$, and a simultaneous eigenstate of $(\hat{\boldsymbol{J}}_a + \hat{\boldsymbol{S}}_b)^2$ with the eigenvalue $3¾\hbar^2$. Now, both $|\Uparrow\rangle_a|\downarrow\rangle_b$ and $|0\rangle_a|\uparrow\rangle_b$ are eigenstates of $\hat{J}_{z,a} + \hat{S}_{z,b}$ with an eigenvalue of $½\hbar$, but neither is an eigenstate of $(\hat{\boldsymbol{J}}_a + \hat{\boldsymbol{S}}_b)^2$, as seen below:

$$\begin{aligned}(\hat{\boldsymbol{J}}_a + \hat{\boldsymbol{S}}_b)^2|\Uparrow\rangle_a|\downarrow\rangle_b &= 2¾\hbar^2|\Uparrow\rangle_a|\downarrow\rangle_b + 2(\hbar/\sqrt{2})(½\hbar)|0\rangle_a|\uparrow\rangle_b + 2(\mathrm{i}\hbar/\sqrt{2})(-½\mathrm{i}\hbar)|0\rangle_a|\uparrow\rangle_b \\ &\quad + 2\hbar(-½\hbar)|\Uparrow\rangle_a|\downarrow\rangle_b = 1¾\hbar^2|\Uparrow\rangle_a|\downarrow\rangle_b + \sqrt{2}\hbar^2|0\rangle_a|\uparrow\rangle_b. \end{aligned} \tag{58}$$

$$\begin{aligned}(\hat{\boldsymbol{J}}_a + \hat{\boldsymbol{S}}_b)^2|0\rangle_a|\uparrow\rangle_b &= 2¾\hbar^2|0\rangle_a|\uparrow\rangle_b + 2(\hbar/\sqrt{2})(½\hbar)(|\Uparrow\rangle + |\Downarrow\rangle)_a|\downarrow\rangle_b \\ &\quad + 2(\mathrm{i}\hbar/\sqrt{2})(½\mathrm{i}\hbar)(-|\Uparrow\rangle + |\Downarrow\rangle)_a|\downarrow\rangle_b \\ &= 2¾\hbar^2|0\rangle_a|\uparrow\rangle_b + \sqrt{2}\hbar^2|\Uparrow\rangle_a|\downarrow\rangle_b. \end{aligned} \tag{59}$$

However, a proper choice of the probability amplitudes $c_1$ and $c_2$ can turn a superposition of $|\Uparrow\rangle_a|\downarrow\rangle_b$ and $|0\rangle_a|\uparrow\rangle_b$ into an eigenstate of total angular momentum, namely,

$$(\hat{\boldsymbol{J}}_a + \hat{\boldsymbol{S}}_b)^2(c_1|\Uparrow\rangle_a|\downarrow\rangle_b + c_2|0\rangle_a|\uparrow\rangle_b) = (1¾c_1 + \sqrt{2}c_2)\hbar^2|\Uparrow\rangle_a|\downarrow\rangle_b + (\sqrt{2}c_1 + 2¾c_2)\hbar^2|0\rangle_a|\uparrow\rangle_b. \tag{60}$$

Setting $c_1 = 1/\sqrt{3}$ and $c_2 = \sqrt{2/3}$, we find $(|\Uparrow\rangle_a|\downarrow\rangle_b + \sqrt{2}|0\rangle_a|\uparrow\rangle_b)/\sqrt{3}$ to be an eigenstate of $(\hat{\boldsymbol{J}}_a + \hat{\boldsymbol{S}}_b)^2$, the eigenvalue being $3¾\hbar^2$. In similar fashion, $(|\Downarrow\rangle_a|\uparrow\rangle_b + \sqrt{2}|0\rangle_a|\downarrow\rangle_b)/\sqrt{3}$ is readily seen to be a simultaneous eigenstate of $(\hat{\boldsymbol{J}}_a + \hat{\boldsymbol{S}}_b)^2$ and $\hat{J}_{z,a} + \hat{S}_{z,b}$, with the respective eigenvalues $3¾\hbar^2$ and $-½\hbar$.

An alternative solution for the coefficients $c_1$ and $c_2$ in Eq.(60) is $c_1 = \sqrt{2/3}$ and $c_2 = -1/\sqrt{3}$, yielding an eigenvalue of $¾\hbar^2 = ½(1 + ½)\hbar^2$ for $(\hat{\boldsymbol{J}}_a + \hat{\boldsymbol{S}}_b)^2$. The corresponding two-particle states $(\sqrt{2}|\Uparrow\rangle_a|\downarrow\rangle_b - |0\rangle_a|\uparrow\rangle_b)/\sqrt{3}$ and $(\sqrt{2}|\Downarrow\rangle_a|\uparrow\rangle_b - |0\rangle_a|\downarrow\rangle_b)/\sqrt{3}$ represent the $|\uparrow\rangle$ and $|\downarrow\rangle$ states of a spin-½ particle.

**12. Concluding remarks**. Following the lines of reasoning pioneered by John S. Bell[5-7] and also advanced by Richard P. Feynman,[4] we have given an account of the incompatibility between quantum mechanics and the existence of local hidden variables. In spite of Einstein's hopes (as expressed at the end of the famous EPR paper[15]) that such hidden variables would someday help to bring about a "complete description of the physical reality," it is now understood that the means for "improving" quantum mechanics (if at all necessary) must be sought elsewhere.

Already in his *Lectures on Physics*[1] (published in 1965) Feynman was dismissive of the EPR paradox. After describing the entangled photon pair experiment (same as the one discussed here in Sec.2), Feynman continues: "*Now many people who learn quantum mechanics in the usual (old fashioned) way find this disturbing. They would like to think that once the photons are emitted it goes along as a wave with a definite character. They would think that since 'any given photon' has some 'amplitude' to be x-polarized or to be y-polarized, there should be some chance of picking it up in either the x- or y-counter and that this chance shouldn't depend on what some other person finds out about a completely different photon. They argue that 'someone else making a measurement shouldn't be able to change the probability that I will find something.' Our quantum mechanics says, however, that by making a measurement on photon number one, you can predict precisely what the polarization of photon number two is going to be when it is detected. This point was never accepted by Einstein, and he worried about it a great deal — it became known as the 'Einstein-Podolsky-Rosen paradox.' But when the situation is described as we have done it here, there doesn't seem to be any paradox at all; it comes out quite naturally that what is measured in one place is correlated with what is measured somewhere else.* ⋯ *[Nature's] way requires a description in terms of interfering amplitudes, one amplitude for each alternative. A measurement of which alternative actually occurs destroys the interference, but if a measurement is not made you cannot still say that 'one alternative or the other is still occurring.'*"[1]